\documentclass{article}

\newlength{\abstractwidth}
\usepackage{cancel}
\usepackage{hyperref}
\usepackage{cite}
\usepackage{color}
\usepackage{xcolor}
\usepackage{graphicx}

\usepackage{amsmath}
\usepackage{amssymb}
\usepackage{latexsym}
 \usepackage{setspace}
 
 \usepackage{caption}
\usepackage{subcaption}

\numberwithin{equation}{section}

\renewcommand{\thefootnote}{\fnsymbol{footnote}}
\renewcommand{\thanks}[1]{\footnote{#1}}
\newcommand{\starttext}{
\setcounter{footnote}{0}
\renewcommand{\thefootnote}{\arabic{footnote}}}
\newcommand{\bea}{\begin{eqnarray}}
\newcommand{\eea}{\end{eqnarray}}
\newcommand{\be}{\begin{eqnarray}}
\newcommand{\ee}{\end{eqnarray}}

\def\ie{\begin{equation}\begin{aligned}}
\def\fe{\end{aligned}\end{equation}}

\def\ie{\begin{equation}\begin{aligned}}
\def\fe{\end{aligned}\end{equation}}
\def\cG{{\cal A}}

\def\cC{{\cal C}}

\def\cG{{\cal G}}

\def\cO{{\cal O}}

\def\cT{{\cal T}}

\def\tr{{\rm tr}}

\usepackage{tikz}
\usetikzlibrary{plotmarks}
\usetikzlibrary{quotes,angles,positioning}
\usetikzlibrary{decorations.pathmorphing}
\usetikzlibrary{decorations.markings}
\usepackage{tikz}
\usetikzlibrary{arrows,shapes}
\usetikzlibrary{trees}
\usetikzlibrary{patterns}
\usetikzlibrary{matrix,arrows} 				
\usetikzlibrary{positioning}				
\usetikzlibrary{calc,through}				
\usetikzlibrary{decorations.pathreplacing}  
\usepackage{pgffor}	
\usetikzlibrary{decorations.pathmorphing}	
\usetikzlibrary{decorations.markings}
\tikzset{
     denom/.style={draw=black,thin},
    scalar/.style={dashed,draw=black, postaction={decorate},
        decoration={markings,mark=at position .55 with {\arrow[draw=black]{>}}}},
}
\tikzstyle{block} = [draw, rectangle,  minimum height=3em, minimum width=6earticlem]
\graphicspath{{./Figs/}} 
\usepackage{tikz}
\usetikzlibrary{decorations.pathreplacing,decorations.markings}
\tikzset{
  on each segment/.style={
    decorate,
    decoration={
      show path construction,
      moveto code={},
      lineto code={
        \path [#1]
        (\tikzinputsegmentfirst) -- (\tikzinputsegmentlast);
      },
      curveto code={
        \path [#1] (\tikzinputsegmentfirst)
        .. controls
        (\tikzinputsegmentsupporta) and (\tikzinputsegmentsupportb)
        ..
        (\tikzinputsegmentlast);
      },
      closepath code={
        \path [#1]
        (\tikzinputsegmentfirst) -- (\tikzinputsegmentlast);
      },
    },
  },
  mid arrow/.style={postaction={decorate,decoration={
        markings,
        mark=at position .55 with {\arrow[#1]{stealth}}
      }}},
}

\begin{document}

\starttext

\setcounter{footnote}{0}

\begin{flushright}
{\small QMUL-PH-22-09}
\end{flushright}

\vskip 0.3in

\begin{center}

{\large \bf Integrated correlators in $\mathcal{N}=4$ super Yang--Mills and periods}

\vskip 0.2in

{Congkao Wen, and Shun-Qing Zhang} 
   
\vskip 0.15in

{\small   Centre for Theoretical Physics, Department of Physics and Astronomy,  }\\ 
{\small Queen Mary University of London,  London, E1 4NS, UK}

\vskip 0.15in

{\tt \small c.wen@qmul.ac.uk, shun-qing.zhang@qmul.ac.uk}

\vskip 0.5in

\begin{abstract}
\vskip 0.1in

We study perturbative aspects of recently proposed integrated four-point correlators in $\mathcal{N}=4$ supersymmetric Yang-Mills with all classical gauge groups using standard Feynman diagram computations. We argue that perturbative contributions of the integrated correlators are given by linear combinations of periods of certain conformal Feynman graphs, which were originally introduced for the construction of perturbative loop integrands of the un-integrated correlator. This observation allows us to evaluate the integrated correlators to high loop orders. We explicitly compute one of the integrated correlators up to four loops in the planar limit, and up to three loops for the other integrated correlator, and find agreement with the results obtained from supersymmetric localisation. The identification between the integrated correlators and certain periods also implies non-trivial relations among these periods, given that one may compute the integrated correlators using localisation. We illustrate this idea by considering one of the integrated correlators at five loops in the planar limit, where the localisation result leads to a prediction for the period of a certain six-loop integral.

   \end{abstract}                                            
   
\end{center}

\baselineskip=15pt
\setcounter{footnote}{0}

\newpage

\setcounter{page}{1}
\tableofcontents
\newpage
\section{Introduction}

The correlation functions of superconformal primary operators  in the stress tensor multiplet of  $\mathcal{N}=4$ super Yang-Mills theory (SYM) have received intensive study both at weak coupling and at strong coupling. Recently, the concept of the integrated correlators was introduced in \cite{Binder:2019jwn}, where it was found that, when integrated over spacetime coordinates with certain integration measures that preserve supersymmetry, the correlators of four superconformal primary operators in $\mathcal{N}=4$ SYM with $SU(N)$ gauge group can be computed using supersymmetric localisation techniques.\footnote{There were two such integrated correlators that have been studied in the literature, and we will refer them as the first integrated correlator and the second integrated correlator, respectively.} This has led to many interesting developments. In particular, the integrated correlators were used as constraints for determining unfixed parameters in the perturbative computation of holographic correlators in $AdS_5 \times S^5$ at supergravity limit and beyond \cite{Binder:2019jwn, Chester:2019pvm, Chester:2020dja}.  Exact results of the integrated correlators with finite complexified Yang-Mills coupling $\tau$ were also obtained, in the large-$N$ expansion \cite{Chester:2019jas, Chester:2020vyz} as well as for arbitrary values of $N$ \cite{Dorigoni:2021bvj, Dorigoni:2021guq}. These exact results of integrated correlators have important applications to the numerical bootstrap of understanding non-BPS operators in $\mathcal{N}=4$ SYM \cite{Chester:2021aun} and to the study of ensemble average of $\mathcal{N}=4$ SYM \cite{Collier:2022emf}. The integrated correlators have been generalised for $\mathcal{N}=4$ SYM with general classical gauge groups, in the large-$N$ expansion \cite{Alday:2021vfb} and for gauge groups with arbitrary ranks and finite coupling $\tau$ \cite{Dorigoni:2022zcr}. One may further extend the integrated correlators for correlation functions with more than four operators; in \cite{Green:2020eyj, Dorigoni:2021rdo}, the integrated  $n$-point maximal $U(1)_Y$-violating correlators were introduced and their implications to the $n$-point maximal $U(1)$-violating superstring amplitudes \cite{Boels:2012zr, Green:2019rhz} in $AdS_5 \times S^5$ were studied.

In this paper, we will study perturbative aspects of integrated correlators using standard Feynman diagram methods. We will compute the integrated correlators order by order in the perturbation expansion using the loop integrands constructed in \cite{Eden:2011we, Eden:2012tu} (see also \cite{Bourjaily:2015bpz, Bourjaily:2016evz} for higher-loop contributions) for the un-integrated correlator. It was observed  in \cite{Eden:2011we} that the integrands of the four-point correlator of superconformal primary operators of  stress-tensor supermultiplets in $\mathcal{N}=4$ SYM has a hidden complete permutation symmetry of external and integration points. This observation has led to very powerful graphical representation of the loop integrands. In particular, the integrands of the four-point correlator at $L$-loop order can be expressed as linear combination of particular graphs, with $(L+4)$ degree-$(-4)$ vertices -- each propagator counts as degree minus one, and each numerator (or inverse propagator) counts as degree plus one. Some of these graphs are simple $4$-regular graphs, but in general they contain numerators. 
These loop integrals have been computed explicitly up to three loops. At one loop \cite{Gonzalez-Rey:1998wyj, Eden:1998hh, Eden:1999kh} and two loops \cite{Eden:2000mv, Bianchi:2000hn}, the resulting correlator is expressed in terms of polylogarithms with transcendental weight two and four, respectively. The three-loop integrals are much harder to evaluate. The correlator at three loops was computed analytically in \cite{Drummond:2013nda}, and the final result involves much more complicated multiple polylogarithms.

To obtain the integrated correlators, in principle one may take these analytical expressions for the un-integrated correlator and then integrate them over spacetime coordinates (more precisely the conformal cross ratios) with the integration measures in the definition of the integrated correlators, as given in \eqref{eq:d2Corr} and \eqref{eq:d4Corr}. However, given the fact that the un-integrated correlator is given by complicated polylogarithms, and even multiple polylogarithms, it is rather challenging to integrate these functions directly with the non-trivial integration measures. Furthermore,  there are no analytical results for the un-integrated correlator beyond three loops, which makes it impossible to study the integrated correlators using Feynman diagram methods at higher loops in this way. 

The observation of this paper is that, instead of taking the analytical results of the un-integrated correlator, it is much more convenient to simply use the loop integrands of the correlator. When integrated with the integration measures that are used in the definition of the integrated correlators, the graphs representing the loop integrands of the un-integrated correlator become precisely the periods of certain  Feynman graphs with vertices of degree-$(-4)$, and such periods have been studied quite extensively in the literature, see for example \cite{Broadhurst:1995km, Schnetz:2008mp, Brown:2009ta, Schnetz:2013hqa, Panzer:2014caa, Schnetz:2016fhy, Panzer:2016snt, Georgoudis:2021onj}. In particular, for the first integrated correlator at $L$ loops, it involves the computation of $(L+1)$-loop periods; for the second integrated correlator at $L$ loops, it is given by a sum of $(L+2)$-loop periods.  Special powerful techniques and packages (such as {\tt HyperInt} \cite{Panzer:2014caa} and {\tt HyperlogProcedures} \cite{HyperlogProcedures}) have been developed for computing these periods, which allow us to evaluate the first integrated correlator up to four loops -- it was computed up to two loops in \cite{Dorigoni:2021guq} -- and up to three loops for the second integrated correlator. We find these results from explicit loop integrals of periods match precisely with the results that are obtained using supersymmetric localisation. 

It should be stressed that the construction of the loop integrands based on the methods of \cite{Eden:2011we, Eden:2012tu}, and hence the periods for the integrated correlators, are general and not specific to the $SU(N)$ gauge group. Especially for the planar sector, which we consider in this paper, the correlator takes a universal form for all classical gauge groups once we use appropriate 't Hooft couplings \cite{Dorigoni:2022zcr}. We therefore compare our Feynman diagram computations with the results obtained from supersymmetric localisation for the integrated correlators in $\mathcal{N}=4$ SYM with general classical gauge groups. The perturbative contribution of the first integrated correlator has in fact been evaluated in \cite{Alday:2021vfb, Dorigoni:2022zcr} using localisation. We will also compute the second integrated correlator for general classical groups using supersymmetric localisation in this paper, for the comparison with the Feynman diagram results.

On the one hand, the agreement between Feynman diagram results and the localisation computation provides important confirmation of the integrated correlators obtained from supersymmetric localisation. The analysis also provides interesting insights of the correlation function in the weak coupling region. In particular, it highlights the simplicity of the integrated correlators. On the other hand, since the integrated correlators can be computed using supersymmetric localisation to arbitrarily high orders, these results from localisation provide very interesting and new relations among the periods associated with these degree-$(-4)$ Feynman graphs that are relevant for the correlator. In particular, when the periods cannot be computed using current techniques, the results of localisation give predictions.  We will illustrate this idea by considering one of the integrated correlators at five loops in the planar limit, the results of localisation lead to a prediction for the analytical expression of a period of a certain six-loop integral. 

The paper is organised as follows.  In section \ref{sec:review}, we will review the integrated four-point correlators in $\mathcal{N}=4$ SYM with general classical gauge groups, and some of the perturbative results obtained from supersymmetric localisation. In section \ref{sec:integrands}, we will review the construction of the loop integrands for the un-integrated four-point correlator. These integrands can be naturally represented in terms of degree-$(-4)$ Feynman graphs. We will then show that once integrated over the integration measures introduced in section \ref{sec:review} for the definition of integrated correlators, they become periods of these degree-$(-4)$ Feynman graphs. In section \ref{sec:evaluation}, we will evaluate all the relevant periods for the first integrated correlator up to four loops in the planar limit, and for the second integrated correlator up to three loops. In both cases, the computation involves periods that are up to five loops. We will also consider the first integrated correlator at five loops in the planar limit. For this case, we are able to compute all the relevant periods except one (they are all six-loop integrals). The known result from localisation then allows us to predict this particular unknown six-loop period. We conclude in section \ref{sec:conclude}, and some technical details of our calculation are described in the appendices.
\section{Integrated correlators in $\mathcal{N}=4$ SYM}
\label{sec:review}

In this section, we will review the definition of integrated four-point correlators in $\mathcal{N}=4$ SYM, and their relations to the localised partition function of $\mathcal{N}=2^{*}$ SYM on $S^4$. We are interested in the correlation function of four superconformal primary operators in the stress-tensor multiplet of $\mathcal{N}=4$ SYM with a gauge group $G_N$, which can be expressed as
\begin{equation} \label{eq:O24}
\langle \cO_2(x_1, Y_1)\dots \cO_2(x_4, Y_4)  \rangle = {1\over x_{12}^4 x_{34}^4} \left[\cT_{G_N,\,\rm free}(U,V;Y_i) + \mathcal{I}_4(U,V; Y_i) \cT_{G_N}(U,V) \right]  \, ,
\end{equation} 
where the superconformal primary operator is defined as $\cO_2(x, Y) :=  \tr(\Phi^{I}(x) \Phi^J(x)) Y_I Y_J$, which has conformal dimension $2$. We have introduced null vector $Y_I$'s ($I=1,2,\cdots,6$) taking care of the $SO(6)$ R-symmetry indices, and the conformal cross ratios, $U,V$ are given by
     \begin{align} \label{eq:UV}
   &U=\frac{x_{12}^2x_{34}^2}{x_{13}^2x_{24}^2}, \; \qquad \quad V=\frac{x_{14}^2x_{23}^2}{x_{13}^2x_{24}^2} \, .
\end{align}
The quantity $\cT_{G_N,\,\rm free}(U,V;Y_i)$ represents the free theory part of the correlator. The non-trivial part of the correlator has been factorised into two pieces: the pre-factor $\mathcal{I}_4(U,V; Y_i)$ is fixed by the superconformal symmetry due to partial non-renormalisation theorem \cite{Eden:2000bk, Nirschl:2004pa} (the expression of $\mathcal{I}_4(U,V; Y_i)$ is given in \eqref{eq:II4}), and $\cT_{G_N}(U,V)$ is the dynamic part of the correlator, which will be the focus of our study. 

We will be interested in the perturbative aspects of the correlator. As we commented in the introduction, in perturbation theory, $\cT_{G_N}(U,V)$ has been computed only up to three loops \cite{Drummond:2013nda}. However the integrands in the planar limit have constructed up to ten loops using very efficient graphic tools \cite{Bourjaily:2016evz}. The non-planar contributions first appear at four loops, and the corresponding integrand is also known \cite{Fleury:2019ydf}.

It was shown in \cite{Binder:2019jwn, Chester:2020dja} that when integrated over suitable integration measures, the correlator can be determined in terms of the partition function of $\mathcal{N}=2^*$ SYM ($\mathcal{N}=4$ SYM with certain mass deformation on the hypermultiplet) on $S^4$, which can be computed using supersymmetric localisation \cite{Pestun:2007rz}. There are two kinds of integrated correlators that have been studied in the literature due to different choices of the integration measures\footnote{Some possible generalisation of these two integrated correlators was suggested in \cite{Naseer:2021cfm}.}. Concretely, they are defined as\footnote{The normalisation of $\mathcal{T}_{G_N}(U,V)$ follows the convention of \cite{Dorigoni:2021guq} and differs from that in \cite{Binder:2019jwn, Chester:2020dja} by a factor of $c^2_{G_N}$, and $c_{G_N}$ is the central charge given in \eqref{eq:c-charge}.}
\begin{align}
\cC_{G_N, 1}(\tau,\bar\tau) 
:=I_2\left[\mathcal{T}_{G_N}(U,V)\right]= - {8\over \pi} \int_0^{\infty} dr \int_0^{\pi} d\theta {r^3 \sin^2(\theta) \over U^2} \cT_{G_N}(U,V) \, ,
\label{eq:d2Corr} 
\end{align}
and 
\begin{align} \label{eq:d4Corr}
\cC_{G_N, 2}(\tau,\bar\tau) :={I}_4 \left[  \cT_{G_N}(U,V) \right]=- {32\over \pi} \int_0^{\infty} dr \int_0^{\pi} d\theta {r^3 \sin^2(\theta) \over U^2}(1+U+V) \bar{D}_{1111}(U, V) \cT_{G_N}(U,V)  \, ,
  \end {align}
where $r$ and $\theta$ are related to cross ratios by $U = 1+r^2 -2r \cos(\theta)$ and $V=r^2$. The function $\bar{D}_{1111}$ is the usual $D$-function that appears in the computation of contact Witten diagrams,  which can be expressed as a one-loop box integral in four dimensions, given by 
\begin{align} \label{eq:boxI}
   \bar{D}_{1111}(U, V)=-\frac{1}{\pi^{2}}\,x^2_{13}x^2_{24}\int \frac{d^4x_5}{x_{15}^2x_{25}^2x_{35}^2x_{45}^2} \,.
    \end{align}
In the notation for the integrated correlators, we have made clear that they are independent of the spacetime coordinates, and they are functions of the (complexified) Yang-Mills coupling   
\begin{equation} \label{eq:coupling}
\tau = \tau_1+i \tau_2 := \frac{\theta}{2\pi}+ i \frac{4\pi}{g_{_{YM}}^2}\,.
\end{equation}
In this paper, we will be mostly concerned with the perturbative contributions, in which case, $\tau_1=0$ (or equivalently the $\theta$ angle vanishes), and the integrated correlators are functions of $\tau_2$ (or $g^2_{_{YM}}$) only.  As we commented earlier, with the choices of the integration measures given in \eqref{eq:d2Corr}  and \eqref{eq:d4Corr}, the integrated correlators are known to be related to the partition function of $\mathcal{N}=2^*$ SYM on $S^4$ through the following relations. For the first integrated correlator, the relation takes the following form,
\begin{align} \label{eq:loc1}
\cC_{G_N,1}(\tau,\bar\tau) =  \frac{1}{4} \Delta_\tau   \partial_m^2 \log Z_{G_N}(\tau,\bar\tau, m)\big{|}_{m=0} \, ,
\end{align}
where the hyperbolic Laplacian is given by $ \Delta_\tau = 4 \tau_2^2 \partial_{\tau}  \partial_{\bar \tau}=\tau_2^2 \left(\partial^2_{\tau_1}+ \partial^2_{\tau_2} \right)$, and $Z_{G_N}(m, \tau,\bar\tau)$ is the partition function of $\mathcal{N}=2^*$ SYM on $S^4$ with $G_N$ gauge group and $m$ is the mass of the hypermultiplet. The second integrated correlator is then given by  
\begin{align} \label{eq:loc2}
\cC_{G_N,2}(\tau,\bar\tau) =  -48 \, \zeta(3) \, c_{G_N} + \partial_m^4 \log Z_{G_N}(m, \tau,\bar\tau)\big{|}_{m=0} \, ,
  \end {align}
where  $c_{G_N}$ is the central charge, 
\begin{equation} \label{eq:c-charge}
c_{SU(N)}=\frac{N^2-1}{4}\,,\qquad 
c_{SO(n)}=\frac{n(n-1)}{8}\,,\qquad 
c_{USp(n)}=\frac{n(n+1)}{8}\, .
\end{equation}
The partition function $Z_{G_N}(m, \tau,\bar\tau)$ can be expressed as a matrix model integral due to supersymmetric localisation \cite{Pestun:2007rz}. Explicitly, it can be expressed as
\ie
 Z_{G_N}(m,\tau,\bar\tau)  = \langle \,   \hat Z_{G_N}^{pert}(m,a) \,  |\hat Z_{G_N}^{inst}  (m ,\tau,a)|^2  \,  \rangle_{G_N}\,,
 \label{partfun}
 \fe
  where we have separated the partition function into the perturbative term $\hat Z_{G_N}^{pert}(m,a)$ and the non-perturbative instanton contribution $\hat Z_{G_N}^{inst}(m ,\tau,a)$. We will omit the instanton contribution, therefore in our consideration $\hat Z_{G_N}^{inst}(m ,\tau,a)=1$. The explicit form of $\hat Z_{G_N}^{pert}(m,a)$ for each classical gauge group $G_N$  and the definition of the expectation value $\langle \cdots \rangle_{G_N}$ can be found in appendix \ref{matrix_model}. Focusing on the perturbative terms, the localisation expressions for integrated correlators reduce to
  \ie \label{eq:loclise3}
  \cC^{pert}_{G_N,1}(\tau_2) &=\frac{1}{4} \tau_2^2\, \partial_{\tau_2}^2 \langle \,   \partial_{m}^2 \hat Z_{G_N}^{pert}(m,a)\big{|}_{m=0} \, \rangle_{G_N}\,, \cr
  \cC^{pert}_{G_N,2}(\tau_2) &=-48\, \zeta(3)\, c_{G_N} + \langle \,   \partial_{m}^4 \hat Z_{G_N}^{pert}(m,a)\big{|}_{m=0} \, \rangle_{G_N} -3\left( \langle \,   \partial_{m}^2 \hat Z_{G_N}^{pert}(m,a)\big{|}_{m=0} \, \rangle_{G_N}\right)^2\, .
  \fe


In the following we will compute the perturbative terms of integrated correlators $\cC^{pert}_{G_N,1}(\tau_2)$ and $  \cC^{pert}_{G_N,2}(\tau_2)$ using the matrix model integrals given in the appendix \ref{matrix_model}. As was found in \cite{Dorigoni:2022zcr}, it is convenient to express the perturbation series in terms of central charge, as given in \eqref{eq:c-charge}, and the 't Hooft coupling
\begin{align} \label{eq:tHooft}
\lambda_{SU(N)}&= g_{_{YM}}^2\, N\,, \qquad\, \lambda_{SO(n)} = g_{_{YM}}^2\,(n-2)\,, \qquad \lambda_{USp(n)} = {g_{_{YM}}^2\,(n+2) \over 2}\, ,
\end{align}
where $\lambda_{SU(N)}$ is the standard 't Hooft coupling for $SU(N)$ gauge group, and the others are the generalisations for other gauge groups \cite{Dorigoni:2022zcr} (see also \cite{Cvitanovic:2008zz}). 

The perturbative expansion for the first integrated correlator was already computed in \cite{Dorigoni:2022zcr}. It was found that $\cC^{pert}_{G_N,1}(\tau_2)$ takes the following universal form for all the gauge groups $G_N$, 
\ie \label{eq:d^2_m}
\cC^{pert}_{G_N,1}(\tau_2) &=4\, c_{G_N} \left[ \frac{3   \, \zeta (3) a_{G_N}   }{2} -\frac{75 \, \zeta (5)a_{G_N}^2}{8} 
+\frac{735 \,\zeta (7) a_{G_N}^3}{16} -\frac{6615  \,\zeta (9)  \left(1 + P_{G_N, 1}\right)  a_{G_N}^4 }  {32} \right. \\
& \qquad\qquad\left. +\frac{114345 \,  \zeta (11) \left(1+  P_{G_N, 2}  \right)a_{G_N}^5  }{128 }+ \mathcal{O}(a_{G_N}^{6}) \right] \, ,
\fe  
where $a_{G_N}={\lambda_{G_N}}/({4 \pi^2})$. We see that the first three perturbative contributions are universal and their dependence on $N$ is contained entirely within $c_{G_N}$ and $a_{G_N}$, therefore the first three loops are all planar, and the non-planar terms only start to enter at four loops.  Furthermore, the planar contribution is universal for all gauge groups. Explicit non-planar  factors, $P_{G_N,i}$ (where $i = L-3$ and $L$ is the loop number), first  enter at four loops and the first two orders for all classical groups are listed below:
\ie \label{eq:PGN}
P_{SU(N), 1}  &= \frac{2}{7N^2} \, , \qquad\qquad \qquad\qquad\quad P_{SU(N), 2} =  {1 \over N^{2}} \, , \cr
P_{SO(n), 1}  &= -\frac{n^2-14 n+32}{14 (n-2)^3}\, , \qquad\qquad P_{SO(n), 2} = -\frac{n^2-14 n+32}{8 (n-2)^3}\,  ,\cr
P_{USp(n), 1}  &= \frac{n^2+14 n+32}{14 (n+2)^3}\ \, ,\qquad \qquad P_{USp(n), 2} = \frac{n^2+14 n+32}{8 (n+2)^3} \, ,
\fe
where for $n=2N$ or $2N+1$ for $SO(n)$, and $n=2N$ for $USp(n)$. It was observed in \cite{Dorigoni:2022zcr} that the expression manifests the relations between the correlators of $SU(N)$ theory and $SU(-N)$ theory, as well as the correlators of $SO(n)$ theory and $USp(-n)$ theory \cite{Mkrtchian:1981bb, Cvitanovic:1982bq}: 
\ie \label{eq:relations}
\cC^{pert}_{SU(N),1}(\tau_2) &= \cC^{pert}_{SU(-N),1}(-\tau_2)\, , \cr
\cC^{pert}_{SO(n),1}(\tau_2) &= \cC^{pert}_{USp(-n),1}(-\tau_2/2)\, .
\fe
It is straightforward to evaluate higher-order terms in perturbative expansion, where one finds similar structures for the integrated correlator, and the relations given in  \eqref{eq:relations} also hold at higher orders.

Similarly, using (\ref{eq:loclise3}) and the matrix model description of the partition function given in appendix \ref{matrix_model}, we have also evaluated the perturbative contributions to the second integrated correlator $\cC_{G_N,2}^{pert}(\tau_2)$, which is given by, 
\begin{align} \label{eq:d^4_m}
\cC_{G_N,2}^{pert}(\tau_2) & = 4\,c_{G_N}
\Bigg[-60 a_{G_N}\zeta(5) + \frac{3a_{G_N}^2(36\zeta(3)^2+175\zeta(7))}{2} -\frac{45 a_{G_N}^3(20\zeta(3)\zeta(5)+49\zeta(9))}{2}\nonumber\\
&+ \frac{45 a_{G_N}^4\,\left(340 \zeta (5)^2+588 \zeta (3) \zeta (7)+1617 \zeta (11)+P_{G_N, 1} \left(840 \zeta (5)^2+1617 \zeta (11)\right)\right)}{16}  \nonumber\\
&-\frac{63a_{G_N}^5\left(1820 \zeta (5) \zeta (7)+1512 \zeta (3) \zeta (9)+4719 \zeta (13)+\frac{21P_{G_N, 1}}{2} (840 \zeta (5) \zeta (7)+144 \zeta (3) \zeta (9)+1573 \zeta (13))\right)}{16} \nonumber\\
&+\mathcal{O}(a_{G_N}^6)\Bigg]\, ,
\end{align}
where the non-planar contribution $P_{G_N, 1}$ is given in \eqref{eq:PGN}. 
We see that $\cC_{G_N,2}^{pert}(\tau_2)$ is considerably more complicated compared to $\cC_{G_N,1}^{pert}(\tau_2)$ (that is also the reason that we do not show the higher-order terms). However, some important features of $\cC_{G_N,1}^{pert}(\tau_2)$ that we commented earlier remain to be true for $\cC_{G_N,2}^{pert}(\tau_2)$. In particular, once again, the planar contribution is universal for gauge groups and the non-planar contributions only start to enter at four loops. The expression given in \eqref{eq:d^4_m} (as well as for the higher-order terms which we did not show explicitly) makes it clear that the relationships \eqref{eq:relations} also hold for $\cC_{G_N,2}^{pert}(\tau_2)$. 

In the following section, we will study these two integrated correlators using Feynman diagram methods. In particular, by using the definitions given in \eqref{eq:d2Corr} and \eqref{eq:d4Corr}, we will argue that applying the loop integrands constructed in \cite{Eden:2011we, Eden:2012tu} using graphical tools, the integrated correlators are given by linear combinations of periods associated with the graphs that represent the loop integrands (and their simple generalisations). By computing these higher-loop periods explicitly, we will show that, up to four loops in the planar limit for the first integrated correlator and up to three loops for the second integrated correlator, the numerical coefficients of the perturbation expansion given in \eqref{eq:d^2_m} and \eqref{eq:d^4_m} agree precisely with the direct loop computations from Feynman diagrams. 
\section{Integrated correlators and Feynman graph periods}
\label{sec:integrands}

In this section, following \cite{Eden:2011we, Eden:2012tu}, we will review the construction of perturbative loop integrands for the four-point correlation function of superconformal primary operators in the stress tensor multiplet of $\mathcal{N}=4$ SYM. It was shown in \cite{Eden:2011we, Eden:2012tu} that due to conformal symmetry and certain hidden permutation symmetry, the Feynman integrals relevant for the correlation function are of very particular forms. At $L$ loops, they are given by the so-called $f^{(L)}$-functions, which can be represented by the so-called $f$-graphs \cite{Bourjaily:2016evz}. We will then argue that these $f^{(L)}$-functions,  when integrated the measures given in \eqref{eq:d2Corr} and \eqref{eq:d4Corr} for the definition of the integrated correlators, are precisely periods of $(L+1)$ loops and $(L+2)$ loops, respectively. Furthermore, these types of Feynman integral periods have been studied in the literature (see e.g. \cite{Broadhurst:1995km, Schnetz:2008mp, Brown:2009ta, Schnetz:2013hqa, Panzer:2014caa, Schnetz:2016fhy, Panzer:2016snt, Georgoudis:2021onj}), and special techniques, especially computer packages, have been developed for their computations. Therefore this observation allows us to compute the integrated correlators to high-loop orders, as we will do in the next section. 

\subsection{Four-point correlator in $\mathcal{N}=4$ SYM and its loop integrands}
\label{sec: loopint}

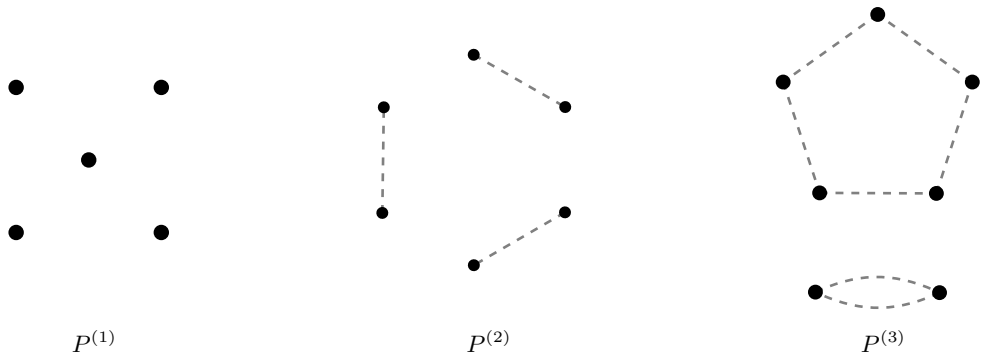
\begin{figure} 
  \centering
  \begin{subfigure}{.3\linewidth}
    \centering
        \begin{tikzpicture}[scale=1.93, line width=1 pt]
            	\filldraw[black] (0.5, 0.5) circle (1.25pt) node[anchor=west]{};
	\filldraw[black] (0.5, -0.5) circle (1.25pt) node[anchor=west]{};
		\filldraw[black] (-0.5, -0.5) circle (1.25pt) node[anchor=west]{};
				\filldraw[black] (-0.5, 0.5) circle (1.25pt) node[anchor=west]{};
	       	\filldraw[black] (0,0) circle (1.25pt) node[anchor=west]{};
\filldraw[white] (0,-1) circle (1.25pt) node[anchor=west]{};
\filldraw[white] (0,1) circle (1.25pt) node[anchor=west]{};           	\end{tikzpicture}
	\caption*{$P^{(1)}$}
     \end{subfigure}
  \hspace{0.5em}
  \begin{subfigure}{.3\linewidth}
    \centering
     \begin{tikzpicture}[scale=1.4, line width=1 pt]
	\begin{scope}[rotate=60]
    \coordinate (A1) at  (-0.8698, 0.4998);
    \coordinate (A2) at  (0.006443, 0.9894);
    \coordinate (A3) at  (0.8655, 0.4998);
    \coordinate (A4) at (0.8698, - 0.5011);
    \coordinate (A5) at (0.002148, - 0.9994);
    \coordinate (A6) at (-0.8655 ,- 0.5011);
	\draw [dashed,gray] (A1)--(A2);
	\draw [dashed,gray] (A3)--(A4);
	\draw [dashed,gray] (A5)--(A6);	
	\filldraw[black] (A1) circle (1.25pt) node[anchor=west]{};
	\filldraw[black] (A2) circle (1.25pt) node[anchor=west]{};
	\filldraw[black] (A3) circle (1.25pt) node[anchor=west]{};
	\filldraw[black] (A4) circle (1.25pt) node[anchor=west]{};
	\filldraw[black] (A5) circle (1.25pt) node[anchor=west]{};
	\filldraw[black] (A6) circle (1.25pt) node[anchor=west]{};
\filldraw[white] (0,-1.2) circle (0.25pt) node[anchor=west]{}; 
\filldraw[white] (1.6,0) circle (0.25pt) node[anchor=west]{}; 
\filldraw[white] (-1.6,0) circle (0.25pt) node[anchor=west]{}; 
	\end{scope}
	\end{tikzpicture}
    \caption*{$P^{(2)}$}
  \end{subfigure}
  \hspace{0.5em}
  \begin{subfigure}{.3\linewidth}
    \centering
    \begin{tikzpicture}[scale=1.32, line width=1 pt]
   \coordinate (A1) at  (-0.5825, - 0.8065);
    \coordinate (A2) at  (0.5931,- 0.8118);
    \coordinate (A3) at  (0.9549, 0.3106);
    \coordinate (A4) at (0.00266, 0.9916);
    \coordinate (A5) at (-0.9496, 0.3106);
 \coordinate (A6) at  (-0.625, - 1.8065);
    \coordinate (A7) at  (0.625,- 1.8118);
	\draw [dashed,gray] (A1)--(A2);
	\draw [dashed,gray] (A2)--(A3);
	\draw [dashed,gray] (A3)--(A4);	
\draw [dashed,gray] (A4)--(A5);
	\draw [dashed,gray] (A5)--(A1);
\draw[dashed,gray] (A6) to [bend left=25] (A7);
\draw[dashed,gray] (A7) to [bend left=25] (A6);
	\filldraw[black] (A1) circle (1.75pt) node[anchor=west]{};
	\filldraw[black] (A2) circle (1.75pt) node[anchor=west]{};
	\filldraw[black] (A3) circle (1.75pt) node[anchor=west]{};
	\filldraw[black] (A4) circle (1.75pt) node[anchor=west]{};
	\filldraw[black] (A5) circle (1.75pt) node[anchor=west]{};
\filldraw[black] (A6) circle (1.75pt) node[anchor=west]{};
	\filldraw[black] (A7) circle (1.75pt) node[anchor=west]{};
	\end{tikzpicture}
    \caption*{$P^{(3)}$}
     \end{subfigure}
  \caption{Here we draw examples of the $P$-graphs for the numerator polynomials $P^{(L)}$ with $L=1,2,3$, they are taken from Fig.1 in \cite{Eden:2012tu}. As one can see that each $P^{(L)}$-graph has $(L+4)$ vertices, and each vertex has degree $(L-1)$.}
  \label{fig:P_graphs}
\end{figure}

\begin{figure} 
  \centering
  \begin{subfigure}{.3\linewidth}
    \centering
     \begin{tikzpicture}[scale=1.75, line width=1 pt]
	\begin{scope}[rotate=270]
    \coordinate (A1) at  (-0.8633,- 0.4959);
    \coordinate (A2) at  (0.8685, -0.5045);
    \coordinate (A3) at  (0.006859,1);
    \coordinate (A4) at (-0.1689,0.1857);
    \coordinate (A5) at (0.1869,0.2199);
    \coordinate (A6) at (0.2941,-0.1787);
    \coordinate (A7) at (0.02401,-0.3673);
    \coordinate (A8) at (-0.2161,- 0.2087);
	\draw    (A1) --(A2);
	\draw    (A1) --(A3);
	\draw    (A1) --(A4);
	\draw    (A1) --(A7);
	\draw    (A1) --(A8);
	\draw    (A2)--(A3);
    \draw    (A2)--(A5);
    \draw    (A2)--(A6);
    \draw    (A2)-- (A7);
	\draw    (A3)--(A4);
	\draw    (A3)--(A5);
	\draw    (A4)--(A5);
	\draw    (A4)--(A6);
	\draw    (A4)--(-0.2161,- 0.2087);
	\draw    (A5)--(A6);
	\draw    (A6)--(A7);
	\draw    (A6)--(A8);
	\draw    (A7)--(A8);
	\filldraw[blue] (A1) circle (1.25pt) node[anchor=west]{};
	\filldraw[blue] (A2) circle (1.25pt) node[anchor=west]{};
	\filldraw[black] (A3) circle (1.25pt) node[anchor=west]{};
	\filldraw[blue] (A4) circle (1.25pt) node[anchor=west]{};
	\filldraw[black] (A5) circle (1.25pt) node[anchor=west]{};
	\filldraw[blue] (A6) circle (1.25pt) node[anchor=west]{};
	\filldraw[black] (A7) circle (1.25pt) node[anchor=west]{};
	\filldraw[black] (A8) circle (1.25pt) node[anchor=west]{};
\draw[dashed,gray] (A1) to [bend left=35] (A6);
\draw[dashed,gray] (A2) to [bend right=35] (A4);
	\end{scope}
	\end{tikzpicture}
	\caption*{$f^{(4)}_1$}
     \end{subfigure}
  \hspace{0.5em}
  \begin{subfigure}{.3\linewidth}
    \centering
    \begin{tikzpicture}[scale=1.75, line width=1 pt]
	\begin{scope}[rotate=150] 
    \coordinate (A1) at  (0.004756, 0.9982);
    \coordinate (A2) at  (-0.8615,-0.4991);
    \coordinate (A3) at  (0.8737,-0.5018);
    \coordinate (A4) at (-0.003265,-0.2932);
    \coordinate (A5) at (0.09833,0.2415);
    \coordinate (A6) at (-0.09684,0.2335);
    \coordinate (A7) at (-0.3936,-0.05794);
    \coordinate (A8) at (0.3978,-0.05794);
	\draw     (A1)--(A2);
	\draw     (A2)--(A3);
	\draw     (A3)--(A1);
	\draw     (A1)--(A5);
	\draw     (A1)--(A6);
	\draw     (A1)--(A7);
	\draw     (A1)--(A8);
	\draw     (A2)--(A4);
	\draw     (A2)--(A7);
	\draw     (A3)--(A4);
	\draw     (A3)--(A8);
	\draw     (A4)--(A5);
	\draw     (A4)--(A6);
	\draw     (A4)--(A7);
	\draw     (A4)--(A8);
	\draw     (A5)--(A6);
	\draw     (A5)--(A8);
	\draw     (A6)--(A7);
	\filldraw[red] (A1) circle (1.25pt) node[anchor=west]{};
	\filldraw[black] (A2) circle (1.25pt) node[anchor=west]{};
	\filldraw[black] (A3) circle (1.25pt) node[anchor=west]{};
	\filldraw[red] (A4) circle (1.25pt) node[anchor=west]{};
	\filldraw[black] (A5) circle (1.25pt) node[anchor=west]{};
	\filldraw[black] (A6) circle (1.25pt) node[anchor=west]{};
	\filldraw[black] (A7) circle (1.25pt) node[anchor=west]{};
	\filldraw[black] (A8) circle (1.25pt) node[anchor=west]{};
\draw[dashed,gray] (A1) to [bend left=5] (A4);
\draw[dashed,gray] (A1) to [bend right=5] (A4);
	\end{scope}
	\end{tikzpicture}
    \caption*{$f^{(4)}_2$}
  \end{subfigure}
  \hspace{0.5em}
  \begin{subfigure}{.3\linewidth}
    \centering
    \begin{tikzpicture}[scale=0.5, line width=1 pt]
	
			\filldraw[black] (0.75,0.75) circle (4pt) node[anchor=west]{};
			\filldraw[black] (0.75,-0.75) circle (4pt) node[anchor=west]{};
			\filldraw[black] (-0.75,0.75) circle (4pt) node[anchor=west]{};
			\filldraw[black] (-0.75,-0.75) circle (4pt) node[anchor=west]{};
			\filldraw[black] (3,0) circle (4pt) node[anchor=west]{};
			\filldraw[black] (-3,0) circle (4pt) node[anchor=west]{};
			\filldraw[black] (0,3) circle (4pt) node[anchor=west]{};
			\filldraw[black] (0,-3) circle (4pt) node[anchor=west]{};
			\draw      (0.75,0.75)--(0.75,-0.75);
			\draw      (0.75,-0.75)--(-0.75,-0.75);
			\draw      (-0.75,-0.75)--(-0.75,0.75);
			\draw      (-0.75,0.75)--(0.75,0.75);
			\draw      (0.75,0.75)--(3,0);
			\draw      (0.75,-0.75)--(3,0);
			\draw      (0.75,0.75)--(0,3);
			\draw      (-0.75,0.75)--(0,3);
			\draw      (-0.75,0.75)--(-3,0);
			\draw      (-0.75,-0.75)--(-3,0);
			\draw      (-0.75,-0.75)--(0,-3);
			\draw      (0.75,-0.75)--(0,-3);
			\draw      (3,0)--(0,3);
			\draw      (0,3)--(-3,0);
			\draw      (-3,0)--(0,-3);
			\draw      (0,-3)--(3,0);
					\end{tikzpicture}
    \caption*{$f^{(4)}_3$}
     \end{subfigure}
  \caption{Here we draw $f$-graphs that contribute the correlator at four loops in the planar limit. They are also given in Fig.8 in \cite{Eden:2012tu}. The solid straight lines denote propagators and the dashed lines are the numerators (or inverse propagators). The blue or red vertex means it has one or two numerator(s) attached, respectively. This ensures that each vertex has degree $(-4)$, if we count each solid straight line as $(-1)$ and dashed line as $(+1)$. }
  \label{fig:4loop_fgraphs}
\end{figure}
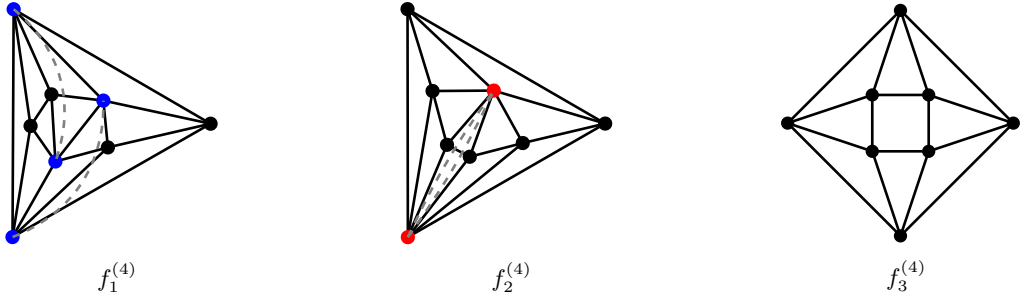

The Feynman integrals that are relevant for the $L$-loop contribution to the  four-point correlator of the superconformal primary operators $\mathcal{O}_2$ operators are the so-called $f^{(L)}$-functions \cite{Eden:2012tu}. In general, $f^{(L)}$-function is given by a linear combination of $f^{(L)}_{\alpha}(x_1, x_2, \ldots, x_{4+L})$ with coefficients that are determined by physical requirements, 
\ie  \label{eq:summ}
f^{(L)}(x_i) = \sum_{\alpha=1}^{n_{L}} c_{\alpha}^{(L)} f^{(L)}_{\alpha}(x_1, x_2, \ldots, x_{4+L}) \, .
\fe
and $f^{(L)}_{\alpha}$ may contain both planar and non-planar topologies. We will only consider the planar ones in this paper. Each function $f^{(L)}_{\alpha}$ is given by
\ie \label{eq:f-function}
f^{(L)}_{\alpha}(x_1, x_2, \ldots, x_{4+L}) = {P^{(L)}_{\alpha}(x_1, x_2, \ldots, x_{4+L}) \over \prod_{1 \leq i < j \leq 4+L} x_{ij}^2 } \, ,
\fe
where the subscript $\alpha$ denotes different planar topologies, and we sum over all $n_L$ number of them, see Table.1 in \cite{Eden:2012tu} for $n_L$ at lower loops. The function $f^{(L)}$ without the subscript $\alpha$ simply means it has only one planar topology, i.e. $n_L=1$.
The numerator $P^{(L)}_{\alpha}$ is a polynomial that is determined by the so-called $P$-graphs. The $P$-graphs are loop-less multigraph with $(4 + L)$ vertices of degree $({L}- 1)$. A line that connects vertices $i, j$ represents a factor $x_{ij}^2$ -- a loop (i.e. a line that connects to the same vertex) is therefore not allowed, it would otherwise lead to a vanishing result, $x_{ii}^2=0$. The function $P^{(L)}_{\alpha}$ is then given by the product of these factors $x_{ij}^2$ associated with a given $P$-graph.  For example see Fig.\ref{fig:P_graphs}, where we give $P$-graphs for $L=1,2,3$. 
It is easy to see that $f^{(L)}_{\alpha}(x_1, x_2, \ldots, x_{4+L})$ has degree-$(-4)$ at each point $x_i$. Furthermore, $f^{(L)}_{\alpha}(x_1, x_2, \ldots, x_{4+L})$ is permutation symmetric due to the hidden permutation symmetry found in \cite{Eden:2011we}. The $f^{(L)}_{\alpha}$-functions can also be represented as graphs: where the solid straight lines denote propagators in \eqref{eq:f-function} and dashed lines denote the numerators, and each vertex has weight $(-4)$ if we count a solid straight line as $(-1)$ and a dashed line $(+1)$. Such graphs are called $f$-graphs \cite{Bourjaily:2016evz}. Examples of such $f^{(L)}$-graphs for $L=4$ is shown in Fig.\ref{fig:4loop_fgraphs} -- they are the loop integrands that contribute to the correlator at four loop in the planar limit, and for the first three loops, they are shown in Fig.\ref{fig:lower_loop_fgraphs}. 

As shown in \cite{Eden:2011we, Eden:2012tu}, these $f^{(L)}$-functions are the building blocks for constructing the $L$-loop integrands for the four-point correlator. In particular, we may write the perturbative expansion of the correlator as
\ie \label{eq:pert}
\langle \cO_2(x_1, Y_1)\dots \cO_2(x_4, Y_4)  \rangle_{pert} = 2 \, c_{G_N}   \sum_{L=1}^{\infty} a_{G_N}^L  \cG_{4}^{(L)}(1,2,3,4) \, , 
\fe
where $c_{G_N}$ is the central charge of gauge group $G_N$ given in \eqref{eq:c-charge}, and $a_{G_N}= \lambda_{G_N}/(4\pi^2)$ with the 't Hooft coupling $\lambda_{G_N}$ defined in \eqref{eq:tHooft}.
The $L$-loop contribution to the correlation function, denoted by $\cG_{4}^{(L)}(1,2,3,4)$, is given by 
\begin{align}\label{corr_loop}
  \cG_{4}^{(L)}(1,2,3,4)=   R(1,2,3,4)   \times  F^{(L)}(x_i)  \,,
\end{align}
where the prefactor $R(1,2,3,4)$ is completely fixed by superconformal symmetries \cite{Eden:2000bk, Nirschl:2004pa}, and is defined as
\begin{align}\label{eq:factorR}
 R(1,2,3,4) &= \frac{Y_{12}Y_{23}Y_{34}Y_{14}}{x^2_{12}x^2_{23}x^2_{34} x^2_{14}}(x_{13}^2 x_{24}^2-x^2_{12} x^2_{34}-x^2_{14} x^2_{23})\nonumber
 \\ &
+\frac{ Y_{12}Y_{13}Y_{24}Y_{34}}{x^2_{12}x^2_{13}x^2_{24} x^2_{34}}(x^2_{14} x^2_{23}-x^2_{12} x^2_{34}-x_{13}^2 x_{24}^2) \nonumber
\\
& +\frac{Y_{13}Y_{14}Y_{23}Y_{24}}{x^2_{13}x^2_{14}x^2_{23} x^2_{24}}(x^2_{12} x^2_{34}-x^2_{14} x^2_{23}-x_{13}^2 x_{24}^2) \nonumber 
 \\ &
+
\frac{Y^2_{12} Y^2_{34}}{x^2_{12}x^2_{34}}   + \frac{Y^2_{13} Y^2_{24}}{x^2_{13}
x^2_{24}} + \frac{Y^2_{14} Y^2_{23}}{x^2_{14}x^2_{23}} \,,
\end{align}
where $Y_{ij}=Y_i \cdot Y_j$, and it is proportional to ${\cal I}_4(U, V, Y_i)$ in \eqref{eq:O24} as, 
\ie \label{eq:II4} {\cal I}_4(U, V, Y_i) = x^2_{13}x^2_{24}\, UV R(1,2,3,4) \, .
\fe
The function $F^{(L)}(x_i)$ is related to $f^{(L)}(x_i)$, as the following
\begin{equation} \label{eq:f_function}
F^{(L)}(x_i)=\frac{x_{12}^2x_{13}^2x_{14}^2x_{23}^2x_{24}^2x_{34}^2}{L!(-4\pi^2)^L} \int d^4x_5 \cdots d^4x_{4+L} \,f^{(L)}(x_i)\, .
\end{equation}
The loop integrands $f^{(L)}(x_i)$ have been computed up to ten loops in the planar limit \cite{Eden:2011we, Eden:2012tu, Bourjaily:2015bpz, Bourjaily:2016evz}, and up to four loops for the non-planar contribution \cite{Fleury:2019ydf}. Finally, comparing \eqref{eq:pert} with \eqref{eq:O24} and using \eqref{eq:II4}, we find the perturbative contribution to the correlator can be expressed as
\ie  \label{eq:TGN}
\cT_{G_N}(U, V) = 2\, c_{G_N} {U \over V} \sum_{L=1}^{\infty} a^L_{G_N} \, x_{13}^2 x_{24}^2\, F^{(L)}(x_i) \, .
\fe 
This is the formula that we will be using for the computation of integrated correlators in the next subsection. 

\subsection{Integrated correlators as Feynman graph periods}

A period \cite{Kontsevich2001} is defined to be the absolutely convergent integral of a rational differential form over a domain given by polynomial inequalities:
\ie \label{eq:period_of_f}
\pi^{-2n}\int_{\Delta} dx_1 \ldots dx_n {P(x_1, \ldots, x_n) \over Q(x_1, \ldots, x_n)}  \, , 
\fe
where the integration domain $\Delta$ is defined by $\{h_i(x_1, \ldots, x_n) \geq 0\}$, and $P, Q, h_i$ are polynomials with rational coefficients. The construction of periods has many fascinating applications to number theory as well as to the computations of Feynman diagrams in quantum field theory. In this subsection, we will argue that the integrated correlators defined in \eqref{eq:d2Corr} and \eqref{eq:d4Corr} using the loop integrands reviewed in subsection \ref{sec: loopint} are precisely periods of Feynman graphs associated with the $f^{(L)}$-functions. In particular, we find the first integrated correlator can be expressed in terms of the periods of $f$-graphs, whereas for the second integrated correlator, it is given by the periods of $\widetilde{f}$-graphs. A $\widetilde{f}$-graph is the generalisation of the $f$-graph by attaching it with an additional one-loop box integral, which will define later with more details.

\subsubsection{First integrated correlator}

We begin with the first correlator given in \eqref{eq:d2Corr},
\begin{align} \label{eq:int1new}
\cC_{G_N,1}(\tau_2) =  
I_2\left[\mathcal{T}_{G_N}(U,V)\right]= - {8\over \pi} \int_0^{\infty} dr \int_0^{\pi} d\theta {r^3 \sin^2(\theta) \over U^2} \cT_{G_N}(U,V) \, ,
\end{align}
then plug in the perturbative contribution of $\mathcal{T}_{G_N}(U,V)$ in \eqref{eq:TGN} to arrive the following expression
\begin{align} 
    I_2\left[\mathcal{T}_{G_N}(U,V)\right]= - {8\over \pi}\,(2c_{G_N})\, \int_0^{\infty} dr \int_0^{\pi} d\theta\, {r^3 \sin^2(\theta) } \frac{1}{UV}\,x^2_{13}x^2_{24} \sum_{L\geq1}a_{G_N}^L F^{(L)}(x_i)\,.
\end{align}
For the later convenience and comparison with the localisation result \eqref{eq:d^2_m}, we will consider the integration acting on $F^{(L)}$ at each loop order, and pull out an overall factor $(4\,c_{G_N})$ as follows
\ie \label{eq:I2_expansion}
I_2\left[\mathcal{T}_{G_N}(U,V)\right] &=4\,c_{G_N} \sum_{L \geq 1} a_{G_N}^L I_2^{\prime} \left[ F^{(L)}(x_i)  \right] \, ,
\fe
and define the $I_2^{\prime}$ integral as
\ie \label{eq:corr_derive}
I^{\prime}_2\left[F^{(L)}(x_i)\right]= - {4\over \pi} \int_0^{\infty} dr \int_0^{\pi} d\theta\, {r^3 \sin^2(\theta) } \frac{1}{UV}\,x^2_{13}x^2_{24} F^{(L)}(x_i)\,. 
\fe
As observed in \cite{Dorigoni:2021guq}, the above integral \eqref{eq:corr_derive} can be viewed as an integration over a four-dimensional vector, $P_V$, with $P_V^2 =V$. Using the relations $U=1+r^2-2r \cos(\theta)$ and $V=r^2$,
we have 
\begin{equation}
    \int d^4 {P}_{_{V}} = 4\pi  \int_0^{\infty}  dr\, r^3 \int_0^{\pi} d\theta  \sin^2(\theta)\,.
\end{equation}
In this form, the integrated correlator can be expressed as
\begin{align} 
    I^{\prime}_2\left[F^{(L)}(x_i)\right]=  -{1\over \pi^2} \int d^4P_V \frac{1}{UV}\,x^2_{13}x^2_{24}\, F^{(L)}(x_i)\,.
\end{align}
Substituting the definition for the cross ratios $U,V$ given in \eqref{eq:UV} in terms of $x_i$'s, we find, 
\begin{align} \label{eq:Ip2}
    I^{\prime}_2\left[F^{(L)}(x_i)\right]&= -{1\over \pi^2} \int d^4P_V \frac{x^6_{13}x^6_{24}}{x^2_{12}x^2_{34}x^2_{14}x^2_{23}}  F^{(L)}(x_i)\,
    \nonumber\\
    &= - {1\over \pi^2 L! (-4\pi^2)^L} \int d^4P_V {x^8_{13}x^8_{24}}  \int d^4x_5 \cdots d^4x_{4+L} \,f^{(L)}(x_i)\, ,
\end{align} 
where we have used the relation \eqref{eq:f_function} to arrive at the final expression. 

The integrated correlator is given by finite conformal integrals, which allow us to fix three points, for the convenience, we choose them to be $(\bold{0}, \bold{1}, \bold{\infty})$. Firstly, we set $x_4$ to be infinity, and the integration \eqref{eq:Ip2} reduces to the following expression, 
\begin{align}
    I^{\prime}_2\left[F^{(L)}(x_i)\right]= -{1\over \pi^2 L! (-4\pi^2)^L} \int d^4P_V {x^8_{13}}  \int d^4x_5 \cdots d^4x_{4+L} \,f^{(L)}(x_i)\,,
\end{align}
and we also observe under the $x_4 \rightarrow \infty$ limit, the cross ratios become
\begin{equation}
    U=\frac{x^2_{12}}{x^2_{13}}, \qquad \quad V=\frac{x^2_{23}}{x^2_{13}}\,.
\end{equation}
We further choose $x_3=\bold{0},\,x_1=\bold{1}$, then $x_2$ is identified with $P_V$. Finally, putting  everything together, we find the integrated  correlator is given by
\begin{align} \label{eq:period_d2mF1}
    I^{\prime}_2\left[F^{(L)}(x_i)\right]&= 
     - {1\over \pi^2 L! (-4\pi^2)^L} \int d^4 x_2 \int d^4x_5 \cdots d^4x_{4+L} \,f^{(L)}(x_i) \Big{|}_{(x_3,x_1,x_4)=(\bold{0},\bold{1},\infty)}\,,
\end{align}
which is exactly the definition of a period with $(x_3,x_1,x_4)=(\bold{0},\bold{1},\infty)$. Therefore, we conclude\footnote{Note the factor $1/(\pi^2)^{L}$ has been absorbed in the definition of periods in (\ref{eq:period_of_f}).}
\begin{align} \label{eq:period_d2mF}
    I^{\prime}_2\left[F^{(L)}(x_i)\right]=-{1\over  L! (-4)^L} \mathcal{P}_{f^{(L)}}\, .
\end{align}
Importantly,  since $f^{(L)}(x_i)$ is permutation invariant, the result is independent of choosing which three points to take special values. 

 Because these graphs are finite and conformal, we may `complete' the graph  by putting back $x_1,x_3,x_4$ in \eqref{eq:period_d2mF1}, in terminology of \cite{Schnetz:2013hqa}. The periods defined above are then associated with $f$-graphs, such as those in Fig.\ref{fig:4loop_fgraphs}.  In the simplest case when the numerator cancels completely some of the denominators, $\mathcal{P}_{f_{\alpha}^{(L)}}$ reduces to the period associated with certain Feynman diagram of the $\phi^4$ theory, and it is then a $4$-regular graph. While for the integrated correlators we consider here, the graphs generally involve numerators, but all the graphs are still restricted to be Feynman graphs with each vertex of degree-$(-4)$.

In summary, we conclude that the first integrated correlator can be expressed as a sum of periods at every loop order,  
\ie 
I_2\left[\mathcal{T}_{G_N}(U,V)\right] &=4\,c_{G_N} \sum_{L \geq 1} a_{G_N}^L I_2^{\prime} \left[ F^{(L)}(x_i)  \right] =-4\, c_{G_N}\sum_{L \geq 1}  { a_{G_N}^L\over L! (-4)^L} \mathcal{P}_{f^{(L)}} 
\cr
&=-4\, c_{G_N} \sum_{L \geq 1} { a_{G_N}^L\over L! (-4)^L} \sum_{\alpha=1}^{n_L} c_{\alpha}^{(L)}\, \mathcal{P}_{f^{(L)}_{\alpha}} 
\, ,
\fe
where we have used \eqref{eq:summ} to arrive at the final expression.

\subsubsection{Second integrated correlator}
\label{sec:secondInt}

Let us now consider the second integrated correlator as defined in (\ref{eq:d4Corr}). We will see that, at $L$ loops, the integrated correlator is expressed as a sum of periods of $(L+2)$ loops. Compared to the first integrated correlator, there is an additional loop integral, this is because of the one-loop box integral $\bar{D}_{1111}(U, V)$ in the integration measure. 

\begin{figure} 
  \centering
  \begin{subfigure}{.3\linewidth}
    \centering\begin{tikzpicture}[scale=1.85, line width=1 pt]
    \begin{scope}[rotate=180]
       \coordinate (A1) at  (0.7666 , 0.003366);
       \coordinate (A2) at  (0.1529 , 0.2965);
       \coordinate (A3) at  (0.0002242 , 0.9773);
       \coordinate (A4) at  (0.4338 , 1.506);
       \coordinate (A5) at  (1.118 , 1.499);
       \coordinate (A6) at  (1.542, 0.9651);
       \coordinate (A7) at  (1.38 , 0.2904);
       \draw (A1) to (A2);
       \draw (A2) to (A3);
       \draw (A3) to (A4);
       \draw (A4) to (A5);
       \draw (A5) to (A6);
       \draw (A6) to (A7);
       \draw (A7) to (A1);
       \draw (A1) to (A3);
       \draw (A1) to (A6);
       \draw (A3) to (A5);
       \draw[line width=1mm,white] (A4) to (A6);
       \draw (A4) to (A6);
        \draw[line width=1mm,white] (A2) to (A7);
       \draw (A2) to (A7);
        \draw[line width=1mm,white] (A2) to (A4);
       \draw (A2) to (A4);
        \draw[line width=1mm,white] (A5) to (A7);
       \draw (A5) to (A7);
\filldraw[black] (A1) circle (1.9pt) node[anchor=west]{};
     \filldraw[gray] (A1) circle (1.5pt) node[anchor=west]{};
	\filldraw[black] (A2) circle (1.5pt) node[anchor=west]{};
	\filldraw[black] (A3) circle (1.5pt) node[anchor=west]{};
	\filldraw[black] (A4) circle (1.5pt) node[anchor=west]{};
	\filldraw[black] (A5) circle (1.5pt) node[anchor=west]{};
    \filldraw[black] (A6) circle (1.5pt) node[anchor=west]{};
	\filldraw[black] (A7) circle (1.5pt) node[anchor=west]{};
       \end{scope}
    \end{tikzpicture}
      \caption*{$\widetilde{f}^{(2,1)}_1$}
     \end{subfigure}
  \hspace{0.5em}
  \begin{subfigure}{.3\linewidth}
    \centering
    \begin{tikzpicture}[scale=2.15, line width=1 pt]
    \begin{scope}[rotate=0]
       \coordinate (A1) at  (0.8503, 1.28 );
       \coordinate (A2) at  (1.596, 0.9594);
       \coordinate (A3) at  (1.606, 0.3086);
       \coordinate (A4) at  (0.8503 , 0.0018);
       \coordinate (A5) at  (0.1004479, 0.3133);
       \coordinate (A6) at  (0.104201, 0.9687);
       \coordinate (A7) at  (0.8503, 0.6433);
       \draw (A1) to (A2);
       \draw (A2) to (A3);
       \draw (A3) to (A4);
       \draw (A4) to (A5);
       \draw (A5) to (A6);
       \draw (A6) to (A1);
       \draw (A1) to (A3);
       \draw (A1) to (A5);
       \draw[line width=1mm,white] (A2) to (A7);
       \draw (A2) to (A7);
       \draw[line width=1mm,white] (A6) to (A7);
       \draw (A6) to (A7);
       \draw (A3) to (A7);
       \draw (A5) to (A7);
       \draw[line width=1mm,white] (A2) to (A4);
       \draw (A2) to (A4);
       \draw[line width=1mm,white] (A4) to (A6);
       \draw (A4) to (A6);
      \filldraw[black] (A1) circle (1.65pt) node[anchor=west]{};
      \filldraw[gray] (A1) circle (1.25pt) node[anchor=west]{};
	\filldraw[black] (A2) circle (1.25pt) node[anchor=west]{};
	\filldraw[black] (A3) circle (1.25pt) node[anchor=west]{};
	\filldraw[black] (A4) circle (1.25pt) node[anchor=west]{};
	\filldraw[black] (A5) circle (1.25pt) node[anchor=west]{};
    \filldraw[black] (A6) circle (1.25pt) node[anchor=west]{};
	\filldraw[black] (A7) circle (1.25pt) node[anchor=west]{};
	\end{scope}
       \end{tikzpicture}
         \caption*{$\widetilde{f}^{(2,2)}_1$}
  \end{subfigure}
  \hspace{0.5em}
  \begin{subfigure}{.3\linewidth}
    \centering
   \begin{tikzpicture}[scale=1.5, line width=1 pt]
    \begin{scope}[rotate=90]
       \coordinate (A1) at  (0.3385 ,1.529);
       \coordinate (A2) at  (1.05, 1.537);
       \coordinate (A3) at  (1.072, -0.31);
       \coordinate (A4) at  (0.3665,-0.31);
       \coordinate (A5) at  (0.002564,0.6071);
       \coordinate (A6) at  (1.416,0.6323);
       \coordinate (A7) at  (1.9,0.6323);
      \draw (A1) to (A2);
     \draw (A2) to (A7);
     \draw (A3) to (A4);
     \draw (A3) to (A7);
     \draw (A4) to (A5);
     \draw (A5) to (A1);
     \draw (A2) to (A3);
     \draw[line width=1mm,white] (A2) to (A6);
       \draw (A2) to (A6);
       \draw[line width=1mm,white] (A3) to (A6);
       \draw (A3) to (A6);
       \draw (A2) to (A5);
       \draw (A3) to (A5);
       \draw[line width=1mm,white] (A1) to (A6);
       \draw (A1) to (A6);
       \draw[line width=1mm,white] (A4) to (A6);
       \draw (A4) to (A6);
       \draw[line width=1mm,white] (A1) to (A7);
       \draw (A1) to (A7);
       \draw[line width=1mm,white] (A4) to (A7);
       \draw (A4) to (A7);
       \draw[line width=1mm,white] (A4) to (A1);
       \draw (A4) to (A1);
       \draw[dashed,gray] (A1) to [bend left=10](A3);
       \draw[dashed,gray] (A2) to [bend left=10](A4);

       \filldraw[blue] (A1) circle (1.75pt) node[anchor=west]{};
	\filldraw[blue] (A2) circle (1.75pt) node[anchor=west]{};
	\filldraw[blue] (A3) circle (1.75pt) node[anchor=west]{};
	\filldraw[blue] (A4) circle (1.75pt) node[anchor=west]{};
	\filldraw[black] (A5) circle (1.75pt) node[anchor=west]{};
    \filldraw[black] (A6) circle (1.75pt) node[anchor=west]{};
    \filldraw[black]  (A7) circle (2.25pt) node[anchor=west]{};
	\filldraw[gray]  (A7) circle (1.75pt) node[anchor=west]{};
	\end{scope}
       \end{tikzpicture}
       \caption*{$\widetilde{f}^{(2,2)}_U$}
     \end{subfigure}
  \caption{Here we draw some examples of $\widetilde{f}$-graphs, $\widetilde{f}^{(2,1)}_1$,$\widetilde{f}^{(2,2)}_1$, $\widetilde{f}^{(2,2)}_U$, which are isomorphic to $f^{(3)}_1$,$f^{(3)}_3$,$f^{(3)}_4$ in Fig.2 in \cite{Eden:2012tu}, respectively.  Other $\widetilde{f}^{(2,k)}_\blacksquare$'s are isomorphic to the graphs shown above or to the three loop planar graph  $f^{(3)}$ in Fig.\ref{fig:lower_loop_fgraphs}. The grey vertices in the graphs denote the point $x_{5+L}$ arising from inserting the one-loop box $\bar{D}_{1111}$ in the definition of the second integrated correlator. We note all the examples of $\widetilde{f}$-graphs are non-planar, even though the corresponding ${f}$-graphs are planar.}
  \label{fig:tilde_fgraphs}
\end{figure}
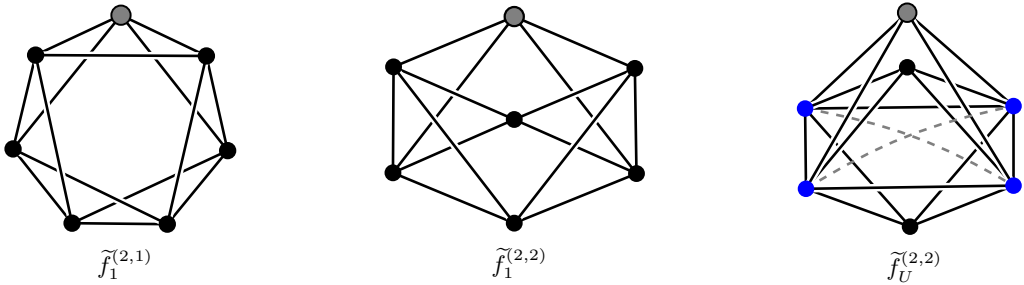

The second integrated correlator is given in (\ref{eq:d4Corr}), which we quote below,  
\begin{align}
&\cC_{G_N,2}(\tau_2) ={I}_4 \left[  \cT_{G_N}(U,V) \right]\nonumber\\
=&- {32\over \pi} \int_0^{\infty} dr \int_0^{\pi} d\theta {r^3 \sin^2(\theta) \over U^2}\, (1+U+V)\bar{D}_{1111}(U,V)\,\cT_{G_N}(U,V) \,.
\end{align}
Using the definition of $\cT_{G_N}(U,V)$ in terms of $F^{(L)}(x_i)$ in (\ref{eq:TGN}), and we arrive at the following expression
\begin{align}
{I}_4 \left[  \cT_{G_N}(U,V) \right]
=- {32\over \pi}\, (2c_{G_N}) \int_0^{\infty} dr \int_0^{\pi} d\theta {r^3 \sin^2(\theta) \over UV} (1+U+V) \bar{D}_{1111}(U,V) \,x^2_{13}x^2_{24}\sum_{L\geq1}a_{G_N}^L F^{(L)}(x_i) \,.
\end{align}
Similarly to (\ref{eq:I2_expansion}) and (\ref{eq:corr_derive}), we define the integral that acts on $F^{(L)}(x_i)$ at each loop order,
\ie 
I_4\left[\mathcal{T}_{G_N}(U,V)\right] &=4\,c_{G_N} \sum_{L \geq 1} a_{G_N}^L I_4^{\prime} \left[ F^{(L)}(x_i)  \right]  \,.
\fe
The $I_4^{\prime}$ integral is defined as
    \begin{align}
        {I}^{\prime}_4 \left[F^{(L)}(x_i)\right]
=\,&-{16\over \pi} \int_0^{\infty} dr \int_0^{\pi} d\theta {r^3 \sin^2(\theta) \over UV} \,x^2_{13}x^2_{24}\, (1+U+V)\bar{D}_{1111}(U,V)\, F^{(L)}(x_i) \nonumber\\
=& \, 4\, {I}^{\prime}_2 \left[\widetilde{F}^{(L)}(x_i) \right] \,,
    \end{align}
where in the last step we have used \eqref{eq:corr_derive} to express 
the $I_4^{\prime}$ integral as the $I_2^{\prime}$ integral 
with a different function in the argument, called $\widetilde{F}^{(L)}(x_i)$, which is defined as
\begin{align}
\widetilde{F}^{(L)}(x_i)= (1+U+V) \bar{D}_{1111}(U,V) F^{(L)}(x_i)\, .  
\end{align}
Furthermore, using the definition of the one-loop box integral  $\bar{D}_{1111}(U,V)$ given in \eqref{eq:boxI}, we can express $\widetilde{F}^{(L)}(x_i)$ as
     \begin{equation} 
\widetilde{F}^{(L)}(x_i)= 4\, \frac{x_{12}^2x_{13}^2x_{14}^2x_{23}^2x_{24}^2x_{34}^2}{L!\,(-4\pi^2)^{L+1}} \int d^4x_5 \cdots d^4x_{4+L} {d^4 x_{5+L}}\, \widetilde{f}^{(L)}(x_i)\, ,
\end{equation}
with 
\ie \label{eq:tildef}
\widetilde{f}^{(L)}(x_i) =\sum_{\blacksquare=1,U,V}\widetilde{f}^{(L)}_{\blacksquare}(x_i) =   \frac{x_{13}^2x_{24}^2 (1+U+V)}{x_{1,5+L}^2x_{2,5+L}^2x_{3,5+L}^2x_{4,5+L}^2} \,f^{(L)}(x_i) \, ,
\fe 
where $\blacksquare$ specifies the three terms, $1$, $U$, and $V$, and we sum over all these terms to ensure $S_4$ permutation symmetry.\footnote{In general, $\widetilde{f}^{(L)}_\blacksquare$ could have another subscript $\alpha$, but since we only consider up to three loops for the second correlator, where $f^{(L)}$ only has one planar topology, we will drop the subscript $\alpha$.}
Since the ${I}^{\prime}_2 \left[\widetilde{F}^{(L)}(x_i) \right]$ integral gives the period of $\widetilde{f}^{(L)}$ as in (\ref{eq:period_d2mF}), we therefore conclude 
\begin{align}\label{eq:period_d4mF}
        {I}^{\prime}_4 \left[F^{(L)}(x_i)\right]
=  4\, {I}^{\prime}_2 \left[\widetilde{F}^{(L)}(x_i) \right]=4\times\, {1 \over L! (-4)^L} \mathcal{P}_{\widetilde{f}^{(L)}} \,.
    \end{align}
Due to the extra one-loop box integral, $\widetilde{f}^{(L)}$ only respects ${S}_{4} \times {S}_{L}$ permutation symmetry instead of full ${S}_{4+L}$ symmetry. The isometry of a graph will then depend on which four vertices are attached to the extra $x_{5+L}$ point, hence may result in different period values. We will call the graphs associated with $\widetilde{f}$-functions as $\widetilde{f}$-graphs, which are $f$-graphs attached with an additional one-loop box (see Fig.\ref{fig:tilde_fgraphs} for some examples). Using the fact that ${f}$-graphs have degree-$(-4)$ at each point $x_i$, \eqref{eq:tildef} implies that $\widetilde{f}$-graphs also have degree-$(-4)$ at each point $x_i$.  Note, due to the additional one-loop box, a $\widetilde{f}$-graph may become non-planar even when the corresponding ${f}$-graph is a planar diagram (as shown in Fig.\ref{fig:tilde_fgraphs}). 

The period of $\widetilde{f}^{(L)}$ needs to be summed over all different isometries, $\widetilde{f}^{(L,k)}$, explicitly
\begin{equation} \label{eq:tilde_f_sum}
    \widetilde{f}^{(L)}(x_i)= \sum_{k=1}^{\widetilde{n}_{L}}\widetilde{f}^{(L,k)}(x_i) \,, 
\end{equation}
where each $\widetilde{f}^{(L,k)}$ respects ${S}_{4} \times {S}_{L}$ permutation symmetry, and $\widetilde{n}_L$ denotes the number of graphs that have distinguished isometries. 
Within each $\widetilde{f}^{(L,k)}$, it contains three terms with different prefactor, $1$, $U$ and $V$. We denote these three terms by $\widetilde{f}^{(L,k)}_1$, $\widetilde{f}^{(L,k)}_U$, and $\widetilde{f}^{(L,k)}_V$ (some examples are given in Fig.\ref{fig:tilde_fgraphs}), therefore, 
\begin{equation}
   \widetilde{f}^{(L,k)}(x_i)=\sum_{\blacksquare=1,U,V}\widetilde{f}^{(L,k)}_{\blacksquare}(x_i) \,.
\end{equation}
To conclude, $\mathcal{P}_{\widetilde{f}^{(L)}}$ can be expressed as, 
\begin{equation} \label{eq:period_d4mF_sum}
    \mathcal{P}_{\widetilde{f}^{(L)}}= \sum_{k=1}^{\widetilde{n}_{L}}\;\sum_{\blacksquare=1,U,V} \mathcal{P}_{\widetilde{f}^{(L,k)}_\blacksquare} \,.
\end{equation}
Here again $\widetilde{n}_{L}$ is the number of the non-isomorphic graphs. 
For example, at two loops, we have two terms (i.e. $\widetilde{n}_{2}=2$), $g \times h(1,2;3,4)+{S}_{4} \times {S}_{2}$ and $g \times [g(1,2,3,4)]^2+{S}_{4}\times {S}_{2}$, which are a one-loop box times a two-loop ladder, and a one-loop box times a square of one-loop boxes, respectively. At three loops, we find $\widetilde{n}_{3}=5$, as we will use later. 

\section{Integrated correlators from Feynman graph periods}
\label{sec:evaluation}

In this section, we will apply the relation between the integrated correlators and periods that we discussed in the previous section to concretely compute the integrated correlators order by order in the perturbative expansion. We will use the Maple package {\tt HyperlogProcedures} developed by Schnetz \cite{HyperlogProcedures} for the evaluations of the periods associated with $f$-graphs and $\widetilde{f}$-graphs to high-loop orders, and find agreement with the perturbation expansion of the integrated correlators obtained using supersymmetric localisation, as given in (\ref{eq:d^2_m}) and (\ref{eq:d^4_m}). 

The identification between the integrated correlators and periods of certain degree-$(-4)$ Feynman graphs also implies interesting relations among these periods. In particular, the sum of these particular periods should produce the results of the integrated correlators that are given by supersymmetric localisation. We will consider the first integrated correlator at five loops in the planar limit, for which, we have computed all the relevant periods, except one. In this case, using the result from supersymmetric localisation, one can predict an analytical expression for the period of a six-loop Feynman graph.

\subsection{First integrated correlator up to four loops}

We begin by considering the first integrated correlator at one and two loops. This was already done in \cite{Dorigoni:2021guq}, and was shown that the results from explicit loop integrals agree with what was obtained from localisation. 
Here we will reproduce these results using the technique of periods (\ref{eq:period_d2mF}). We will then present new results of the integrated correlator at three and four loops. 

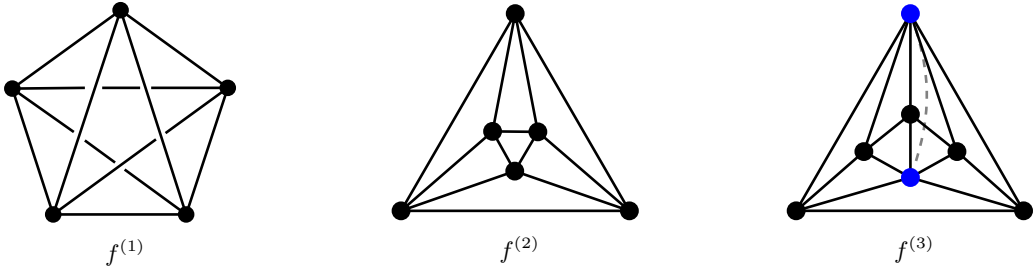
\begin{figure} 
  \centering
  \begin{subfigure}{.3\linewidth}
    \centering
    \begin{tikzpicture}[scale=1.5, line width=1 pt]
     \coordinate (A) at (-0.5924, - 0.8094);
    \coordinate (B) at (0.5858,- 0.8094);
    \coordinate (C) at (0.9547, 0.3133);
    \coordinate (D) at (0.004896, 1.002);
    \coordinate (E) at (-0.9547,0.3068);
    \coordinate (F) at (-0.3117 , 0.3133);
    \coordinate (G) at (-0.1583 , 0.3133);
    \coordinate (H) at (0.1779 , 0.3133);
    \coordinate (I) at (0.2758 , 0.3133);
    \coordinate (J) at (0.4031 , -0.08813);
    \coordinate (K) at (0.3182, -0.1436);
    \coordinate (L) at (0.0408, -0.4112);
    \coordinate (M) at (-0.04406, -0.3492);
    \coordinate (N) at (-0.3117, -0.1534);
    \coordinate (O) at (-0.4096,-0.07833);
    \filldraw[black] (A) circle (1.75pt) node[anchor=west]{};
	\filldraw[black] (B) circle (1.75pt) node[anchor=west]{};
	\filldraw[black] (C) circle (1.75pt) node[anchor=west]{};
	\filldraw[black] (D) circle (1.75pt) node[anchor=west]{};
	\filldraw[black] (E) circle (1.75pt) node[anchor=west]{};
    \draw   (A)--(B);
    \draw   (B)--(C);
    \draw   (C)--(D);
    \draw   (D)--(E);
    \draw   (E)--(A);
    \draw   (E)--(F);
    \draw   (G)--(H);
    \draw   (I)--(C);
    \draw   (C)--(J);
    \draw   (A)--(D);
    \draw   (A)--(K);
    \draw   (B)--(D);
    \draw   (B)--(L);
    \draw   (M)--(N);
    \draw   (O)--(E);
\end{tikzpicture}
	\caption*{$f^{(1)}$}
     \end{subfigure}
  \hspace{0.5em}
  \begin{subfigure}{.3\linewidth}
    \centering
    \begin{tikzpicture}[scale=1.75, line width=1 pt]
\begin{scope}[rotate=0]
 \coordinate (A1) at (0,0.9997);
    \coordinate (A2) at (-0.8672,- 0.4995);
    \coordinate (A3) at (0.8672,-0.4995);
    \coordinate (A4) at (-0.003511,- 0.1975);
    \coordinate (A5) at (0.172, 0.1009);
    \coordinate (A6) at (-0.172,0.1044);
    \filldraw[black] (A1) circle (1.75pt) node[anchor=west]{};
	\filldraw[black] (A2) circle (1.75pt) node[anchor=west]{};
	\filldraw[black] (A3) circle (1.75pt) node[anchor=west]{};
	\filldraw[black] (A4) circle (1.75pt) node[anchor=west]{};
	\filldraw[black] (A5) circle (1.75pt) node[anchor=west]{};
	\filldraw[black] (A6) circle (1.75pt) node[anchor=west]{};
    \draw   (A1)--(A2);
    \draw   (A2)--(A3);
    \draw   (A1)--(A3);
    \draw   (A1)--(A5);
    \draw   (A1)--(A6);
    \draw   (A2)--(A4);
    \draw   (A2)--(A6);
    \draw   (A4)--(A3);
    \draw   (A5)--(A3);
    \draw   (A4)--(A5);
    \draw   (A4)--(A6);
    \draw   (A5)--(A6);
    \end{scope}
\end{tikzpicture}
    \caption*{$f^{(2)}$}
  \end{subfigure}
  \hspace{0.5em}
  \begin{subfigure}{.3\linewidth}
    \centering
     \begin{tikzpicture} [scale=1.75, line width=1 pt]
     \coordinate (A1) at (0.005035, 0.9996);
    \coordinate (A2) at (0.005035, -0.2463);
    \coordinate (A3) at (-0.8629, -0.4992);
    \coordinate (A4) at (0.8645, -0.4992);
    \coordinate (A5) at (-0.3485, -0.0491);
    \coordinate (A6) at (0.3586, -0.0491);
    \coordinate (A7) at (0.005035, 0.2349);
    \draw[dashed,gray] (A1) to [bend left=20] (A2);
    \filldraw[black] (A3) circle (1.75pt) node[anchor=west]{};
	\filldraw[black] (A4) circle (1.75pt) node[anchor=west]{};
	\filldraw[black] (A5) circle (1.75pt) node[anchor=west]{};
	\filldraw[black] (A6) circle (1.75pt) node[anchor=west]{};
	\filldraw[black] (A7) circle (1.75pt) node[anchor=west]{};
	\draw   (A1)--(A3);
    \draw   (A1)--(A4);
    \draw   (A1)--(A5);
    \draw   (A1)--(A6);
    \draw   (A1)--(A7);
    \draw   (A2)--(A3);
    \draw   (A2)--(A4);
     \draw   (A2)--(A5);
    \draw   (A2)--(A6);
    \draw   (A2)--(A7);
    \draw   (A3)--(A4);
    \draw   (A3)--(A5);
    \draw   (A4)--(A6);
    \draw   (A5)--(A7);
    \draw   (A6)--(A7);
    \filldraw[blue] (A1) circle (1.75pt) node[anchor=west]{};
	\filldraw[blue] (A2) circle (1.75pt) node[anchor=west]{};
   \end{tikzpicture}
     \caption*{$f^{(3)}$}
     \end{subfigure}
  \caption{Here we draw the planar $f$-graphs up to three loops. (They have been given in Fig.2 in \cite{Eden:2012tu}). }
  \label{fig:lower_loop_fgraphs}
\end{figure}

Recall that in general the first integrated correlator can be expressed as a sum of periods, 
\ie \label{eq:period_tech}
 I_2\left[\mathcal{T}_{G_N}(U,V)\right]=4\, c_{G_N} \sum_{L \geq 1} a_{G_N}^L I_2^{\prime} \left[ F^{(L)}(x_i)  \right] =-4\,c_{G_N}\sum_{L \geq 1}  \frac{a_{G_N}^L}{ L! (-4)^L} \sum_{\alpha=1}^{n_L} c_{\alpha}^{(L)}\, \mathcal{P}_{f^{(L)}_{\alpha}} 
\, .
\fe
At one and two loops, as shown in Fig.\ref{fig:P_graphs}, the $P^{(L)}$-graphs with $L=1,2$, are unique. Therefore, the associated loop integrands $f^{(1)}, f^{(2)}$ are also unique, as we show in  Fig.\ref{fig:lower_loop_fgraphs}. Using (\ref{eq:period_tech}), we have 
\begin{equation}
    I^{\prime}_2\left[{F^{(1)}}(x_i)\right]=-\frac{1}{1!(-4)^1} \times \mathcal{P}_{f^{(1)}} \,,
\end{equation}
where the $f^{(1)}$-function, see Fig.\ref{fig:lower_loop_fgraphs}, is given by
\begin{align} 
f^{(1)}(x_i)=c^{(1)} \, \frac{P^{(1)}(x_1, \ldots, x_5)}{\prod_{1\le i< j \le 5} x_{ij}^2}\,, \qquad {\rm with} \quad   \quad c^{(1)}=1\,, \qquad P^{(1)}(x_1, \ldots, x_5)=1\, ,
\end{align}
Using the period of $f^{(1)}(x_i)$, which is well-known, 
\begin{equation}
     \mathcal{P}_{f^{(1)}}= 6\zeta(3)\,,
\end{equation}
we arrive at 
\ie \label{eq:FF1}
 I^{\prime}_2\left[{F^{(1)}}(x_i)\right]=\frac{3\zeta(3)}{2}\, .
 \fe

Similarly, at two loops (i.e. $L=2$), 
the $f^{(2)}$-function, see Fig.\ref{fig:lower_loop_fgraphs}, is given by
\begin{align} 
f^{(2)}(x_i)&=c^{(2)} \frac{P^{(2)}(x_1, \dots, x_6)}{\prod_{1\le i< j \le 6} x_{ij}^2}\,, 
\end{align}
where the coefficient $c^{(2)}=1$ and the numerator $P^{(2)}$ is given by
\begin{equation}
    P^{(2)}(x_1, \dots, x_6)=\left( \frac{1}{48}x_{12}^2 x_{34}^2 x_{56}^2\right)+S_6\,.
\end{equation}
A few comments are in order regarding the numerator $P^{(2)}$, which will also be useful for higher-loop computations. Here $S_{4+L}$ (in this case $L=2$) denotes total permutations of ($x_1, \cdots, x_{4+L}$) labels. The factor $48$ ensures the each term in the sum appears with a unit weight, i.e. it mods out the over-counting of $S_{4+L}$ permutations. There are $6!$ terms in ${S}_6$ permutations, while they all have the same value when taking periods. So the $\mathcal{P}_{f^{(2)}}$ is simply given by $6!$ (and divided by 48) times the period of a single term. For example, we take the first term in $P^{(2)}$, which is $x_{12}^2 x_{34}^2 x_{56}^2$, divided by numerator $\prod_{1\le i< j \le 6} x_{ij}^2$, and this single term gives a period with value of $20\zeta(5)$.
So the period of $f^{(2)}$ is given by
\begin{equation} 
     \mathcal{P}_{f^{(2)}}=6!\times\frac{1}{48}\times 20 \zeta(5)\,.
\end{equation}
Put all the factors together, we finally obtain, 
\begin{equation} \label{eq:FF2}
   I^{\prime}_2\left[{F^{(2)}}(x_i) \right]=      -{1\over 2!\, (-4)^2}\, c^{(2)}\, \mathcal{P}_{f^{(2)}} =-\frac{75\zeta(5)}{8}\,. 
\end{equation}
We see that the results of $L=1,2$ cases given in \eqref{eq:FF1} and \eqref{eq:FF2} reproduce the computation of \cite{Dorigoni:2021guq} and match precisely with the first two orders of the localisation result given in \eqref{eq:d^2_m}. The same methods apply to higher-loop terms, and below we will consider three- and four-loop cases.

At three loops, it was shown in \cite{Eden:2011we} that even though one may be able to draw graphs with non-planar topologies at this order, only the planar diagram (and there is a single such planar diagram) can contribute to the four-point correlator. Therefore, just as the one- and two-loop cases, there is a unique integrand at this order, as shown in Fig.\ref{fig:lower_loop_fgraphs}, and it is given by
\begin{align} \label{eq:f3_period}
f^{(3)}(x_i)&=c^{(3)} \frac{P^{(3)}(x_1, \dots, x_7)}{\prod_{1\le i< j \le 7} x_{ij}^2} \,, 
\end{align}
with coefficient $c^{(3)}=1$ and the numerator given by
\ie
P^{(3)}(x_1, \dots, x_7)=\left(\frac{1}{20} x_{12}^4 x_{34}^2 x_{45}^2 x_{56}^2 x_{67}^2 x_{37}^2\right) 
+{S}_7 \,.
\fe
There are $7!$ terms in ${S}_7$ permutations (the factor $20$ again ensures the unit weight of each term in ${S}_7$ permutations), while they all have the same value when taking periods. So the $\mathcal{P}_{f^{(3)}}$ is simply given by $7!$ (and divided by 20) times the period of a single term, which is given by
\begin{equation}
     \mathcal{P}_{f^{(3)}}=7!\times\frac{1}{20} \times 70 \zeta(7)\,,
\end{equation} 
where we have used the Maple program {\tt HyperLogProcedures} to the evaluate this period\footnote{Recall that the periods associated with the $L$-loop contributions to the first integrated correlator are $(L+1)$-loop integrals. So in this case with $L=3$, the period is a four-loop integral.}. Together, we find
\begin{equation} \label{eq:3loops}
    I^{\prime}_2\left[{F^{(3)}}(x_i)\right]=-\frac{1}{3!\,(-4)^3} \times c^{(3)}\, \mathcal{P}_{f^{(3)}}=\frac{735\zeta(7)}{16} ,
\end{equation}
which agrees with the localisation result (\ref{eq:d^2_m}). 

We would like to remark that the three-loop integration over points $x_5, x_6, x_7$ of $f^{(3)}$ leads to the three-loop contribution to the correlator \cite{Drummond:2013nda}, where the answer was found to be expressed in terms of rather complicated multiple polylogarithms. As we showed above, with one additional integral with the measure given in \eqref{eq:int1new}, the result actually simplifies dramatically and reduces to simply some rational number times $\zeta(7)$, as given in \eqref{eq:3loops}.  This example shows clearly the simplicity of the integrated correlator.

Starting at $L=4$, there are non-trivial non-planar contributions \cite{Eden:2012tu, Fleury:2019ydf}. We will only focus on the planar contribution here. At this order, there are three planar $f$-graphs (see Fig.\ref{fig:4loop_fgraphs}), explicitly they are expressed as
\begin{align}
    f^{(4)} (x_i)=\sum_{\alpha=1}^3 c_{\alpha}^{(4)} f^{(4)}_{\alpha}(x_1,\cdots,x_8)=\sum_{\alpha=1}^3\,c_{\alpha}^{(4)}\frac{P_{\alpha}^{(4)}(x_1, \dots, x_8)}{\prod_{1\le i< j \le 8} x_{ij}^2}\,,
\end{align}
where $c^{(4)}_1=c^{(4)}_2=-c^{(4)}_3=1$, and the numerator $P_{\alpha}^{(4)}$'s are given by
\begin{align}  \label{P-4loop-p}
P^{(4)}_1(x_1, \dots, x_8)&=\left(\frac{1}{8}\,x_{12}^2 x_{13}^2 x_{16}^2  x_{24}^2 x_{27}^2 x_{34}^2 x_{38}^2 x_{45}^2 x_{56}^4 x_{78}^4\right)+\text{$S_8$}\,,\nonumber \\ 
P^{(4)}_2(x_1, \dots, x_8)&=\left(\frac{1}{24}\, x_{12}^2 x_{13}^2 x_{16}^2 x_{23}^2 x_{25}^2 x_{34}^2 x_{45}^2 x_{46}^2 x_{56}^2 x_{78}^6\right)+\text{$S_8$}\,, \nonumber\\
P^{(4)}_3(x_1, \dots, x_8)&=\left(\frac{1}{16}\, x_{12}^2 x_{15}^2 x_{18}^2 x_{23}^2 x_{26}^2 x_{34}^2
   x_{37}^2 x_{45}^2 x_{48}^2 x_{56}^2 x_{67}^2 x_{78}^2\right)+\text{$S_8$}\,.
 \end{align}
The first integrated correlator at four loops (the planar sector) is then given by
\begin{equation} \label{eq:4loops}
    I^{\prime}_2\left[{F^{(4)}} (x_i)\right]=-\frac{1}{4!\,(-4)^4} \times \left( \mathcal{P}_{f^{(4)}_1} +\mathcal{P}_{f^{(4)}_2} -\mathcal{P}_{f^{(4)}_3}\right)=-\frac{6615\zeta(9)}{32} \, , 
\end{equation}
where we have used the results of each period of $f^{(4)}_\alpha$
\begin{align}
     &\mathcal{P}_{f^{(4)}_1}= 8!\times \frac{1}{8} \times 252\zeta(9) \,, \nonumber \\
    & \mathcal{P}_{f^{(4)}_2}= 8!\times \;\frac{1}{24} \times 252\zeta(9) \,, \nonumber \\
     &\mathcal{P}_{f^{(4)}_3}= 8!\times \frac{1}{16} \times 168\zeta(9) \nonumber \,.
\end{align}
We have again utilised Maple package {\tt HyperLogProcedures} and the ${S}_8$ permutation symmetry of $f^{(4)}$-graph periods for the computation. The Feynman diagram result \eqref{eq:4loops} once again agrees with the localisation computation given in (\ref{eq:d^2_m}) for the planar part at the order $a_{G_N}^4$.

\subsection{First integrated correlator at five loops and relations of periods} 

As we anticipated, beyond four loops, we have not computed all the periods that are relevant for the first integrated correlator and cannot compare with the supersymmetric localisation results. We will however take a different point of view by considering the localisation results as constraints on these higher-loop Feynman periods. This then leads to non-trivial relations among these periods. In the four-loop example we considered in the previous subsection, one may consider \eqref{eq:4loops} as the required relationship of the five-loop periods for the graphs in Fig.\ref{fig:4loop_fgraphs}.  We will now apply this consideration to the five-loop integrated correlator, which will lead to prediction for a particular six-loop period. 

At five loops, there are seven planar $f$-graphs that contribute to the four-point correlator in the planar limit \cite{Eden:2012tu}, which can be written as
\begin{align}
    f^{(5)} (x_i)=\sum_{\alpha=1}^7 c_{\alpha}^{(5)} f^{(5)}_{\alpha}(x_1,\cdots,x_9)\, ,
\end{align}
with the coefficients determined in \cite{Eden:2012tu}, $c^{(5)}_2=c^{(5)}_3=c^{(5)}_4=c^{(5)}_6=c^{(5)}_7=1$ and $c^{(5)}_1=c^{(5)}_5=-1$.
The explicit forms of $f^{(5)}_{\alpha}$ are given in equations (6.2) and (6.5) of the paper \cite{Eden:2012tu} in terms of $P$-polynomials. They are also shown in Figure 9 in \cite{Eden:2012tu}. 

\begin{figure}
    \centering
    \begin{tikzpicture}[scale=2.5, line width=1 pt]
    \begin{scope}[rotate=90]
       \coordinate (A1) at  (0.552, 0.108);
       \coordinate (A2) at  (0.5575,   1.4885);
       \coordinate (A3) at  (1.7469, 0.7947);
       \coordinate (A4) at  (0.8363, 0.5841);
       \coordinate (A5) at  (0.8301,   0.9991);
       \coordinate (A6) at  (1.1832,   0.7885);
       \coordinate (A7) at  (0.6566,   0.8009);
      \coordinate (A8) at  (1.1027,   1.0363);
       \coordinate (A9) at  (1.09, 0.5407);
      \draw (A1) to (A2);
     \draw (A1) to (A3);
     \draw (A1) to (A4);
     \draw (A1) to (A7);
     \draw (A1) to (A9);
     \draw (A2) to (A3);
     \draw (A2) to (A5);
     \draw (A2) to (A7);
     \draw (A2) to (A8);
      \draw (A3) to (A6);
      \draw (A3) to (A8);
      \draw (A3) to (A9);
       \draw (A4) to (A5);
      \draw (A4) to (A6);
      \draw (A4) to (A7);
      \draw (A4) to (A9);
       \draw (A5) to (A6);
      \draw (A5) to (A7);
      \draw (A5) to (A8);
      \draw (A6) to (A8);
      \draw (A6) to (A9);
     \draw[dashed,gray] (A1) to [bend left=5] (A5);
     \draw[dashed,gray] (A2) to [bend left=5] (A6);
     \draw[dashed,gray] (A3) to [bend left=5] (A4);
         \filldraw[blue] (A1) circle (1.5pt) node[anchor=west]{};  
	\filldraw[blue] (A2) circle (1.5pt) node[anchor=west]{};
	\filldraw[blue] (A3) circle (1.5pt) node[anchor=west]{};
	\filldraw[blue] (A4) circle (1.5pt) node[anchor=west]{};
	\filldraw[blue] (A5) circle (1.5pt) node[anchor=west]{};
    \filldraw[blue] (A6) circle (1.5pt) node[anchor=west]{};
	\filldraw[black]  (A7) circle (1.5pt) node[anchor=west]{};
	\filldraw[black]  (A8) circle (1.5pt) node[anchor=west]{};
	\filldraw[black]  (A9) circle (1.5pt) node[anchor=west]{};
	\end{scope}
	       \end{tikzpicture} 
       \caption{The graph for $f^{(5)}_4$, whose period is predicted by supersymmetric localisation in (\ref{eq:f5_prediction}).}
       \label{fig:f5period}
\end{figure}
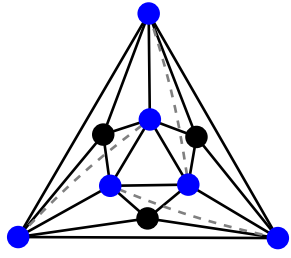

Using {\tt HyperlogProcedures}, we have evaluated all the periods for $f^{(5)}_{\alpha}$,  except for $f^{(5)}_4$, which is shown in Fig.\ref{fig:f5period}. The results of these periods that we have evaluated are listed below: 
\begin{align}
     &\mathcal{P}_{f^{(5)}_1}= 9!\times \frac{1}{2} \times \left[ -40\zeta(3)^2\zeta(5) + 4240\zeta(11) + \frac{8\pi^6}{63}\zeta(5) - \frac{8\pi^4}{5}\zeta(7) - 360\pi^2\zeta(9) - 48\zeta(5, 3, 3)\right] \, , \\
     &\mathcal{P}_{f^{(5)}_2}= 9!\times \frac{1}{4} \times 924\zeta(11)   \, , \\
     &\mathcal{P}_{f^{(5)}_3}= 9!\times \frac{1}{4} \times 924\zeta(11) \, , \\
     &\mathcal{P}_{f^{(5)}_5}= 9!\times \frac{1}{8} \times \Big[ 320\zeta(3)^2\zeta(5) + 800\zeta(5)^2 - 29300\zeta(11)   - \frac{64\pi^6}{63}\zeta(5) + \frac{64\pi^4}{5}\zeta(7) + 2880\pi^2\zeta(9) + 384\zeta(5, 3, 3)\Big]  \, , \\
     &\mathcal{P}_{f^{(5)}_6}= 9!\times \frac{1}{28} \times 924\zeta(11) \, ,  \\
     &\mathcal{P}_{f^{(5)}_7}= 9!\times \frac{1}{12} \times 400\zeta(5)^2   \,.
\end{align}
The multiple zeta value is defined as
\begin{equation}
    \zeta(n_d,\cdots,n_1)=\sum_{k_d>\cdots>k_1\geq 1} \frac{1}{k_d^{n_d}\,\cdots\,k_1^{n_1}}\,  ,\quad\quad n_d\geq 2\,. 
\end{equation}
In order to match with the localisation result given in \eqref{eq:d^2_m} at order $a_{G_N}^5$, it requires the following relations among the periods of $f^{(5)}_{\alpha}$ to hold 
\begin{equation} \label{eq:relation}
    I^{\prime}_2\left[{F^{(5)}}\right]=-\frac{1}{5!\,(-4)^5} \times \left(-\mathcal{P}_{f^{(5)}_1} +  \mathcal{P}_{f^{(5)}_2} +\mathcal{P}_{f^{(5)}_3}+\mathcal{P}_{f^{(5)}_4}  -\mathcal{P}_{f^{(5)}_5}+\mathcal{P}_{f^{(5)}_6}+\mathcal{P}_{f^{(5)}_7}   \right)=\frac{114345}{128} \zeta(11)\, .
\end{equation}
Knowing $\mathcal{P}_{f^{(5)}_{\alpha}}$ for all the $\alpha$'s except $\alpha=4$ (as given in Fig.\ref{fig:f5period}), the above relation allows us to determine the period $\mathcal{P}_{f^{(5)}_4}$, which we find to be
\begin{align} \label{eq:f5_prediction}
    \mathcal{P}_{f^{(5)}_4}=& \, 9!\times \frac{1}{6} \times \Big[120\zeta(3)^2\zeta(5) + 400\zeta(5)^2 - 10410\zeta(11) \cr
    & \qquad \qquad - \frac{8\pi^6}{21}\zeta(5) + \frac{24\pi^4}{5}\zeta(7) + 1080\pi^2\zeta(9) + 144\zeta(5, 3, 3) \Big]\,,
\end{align}
where the prefactor $9!\times \frac{1}{6}$ is some combinatorics factors associated with the permutation symmetry of the integrand, and the numerical value of $\mathcal{P}_{f^{(5)}_4}$ is $9!\times \frac{1}{6}\times (967.13267\cdots)$.

\subsection{Second integrated correlator up to three loops}

In this subsection, we will consider the second integrated correlator.  As we argued in the section \ref{sec:secondInt}, the $L$-loop contribution of the second integrated correlator can be expressed in terms of periods of $(L+2)$ loops. We will compute all the second integrated correlators up to three loops. 

Let us begin by considering the integrated correlator at one loop. Using (\ref{eq:period_d4mF}) and (\ref{eq:period_d4mF_sum}), we have  
\begin{align}
    I^{\prime}_4\left[{F^{(1)}}(x_i)\right] =4\times \frac{1}{(-4)^1} \times \mathcal{P}_{\widetilde{f}^{(1,1)}}\,,
    \end{align}
where
\begin{align}
    \widetilde{f}^{(1,1)}(x_i) =\sum_{\blacksquare = 1, U, V}\widetilde{f}^{(1,1)}_{\blacksquare}(x_i)
    = \frac{x_{13}^2x_{24}^2\,(1+U+V)}{\prod_{1\leq i < j\leq 4} x_{ij}^2}\, g\times  g^{(5)}_{(1,2,3,4)}\,.
\end{align}The function $g^{(5)}_{(1,2,3,4)}$ is the integrand of the one-loop box
\begin{equation}
g^{(5)}_{(1,2,3,4)}=\frac{1} {{x_{1,5}^2x_{2,5}^2x_{3,5}^2x_{4,5}^2}}\,,
\end{equation}
and we define a short-hand notation, $g\times[\cdots]$, which is the product of a one-loop box $g$ and an $L$-loop integrand,
\begin{equation}
g\times[\cdots]:=\frac{1} {{x_{1,5+L}^2x_{2,5+L}^2x_{3,5+L}^2x_{4,5+L}^2}}\times [\cdots]\,,
\end{equation}
where $[\cdots]$ is the integrand of any $L$-loop integral.
The period of $\widetilde{f}^{(1,1)}$ can be evaluated straightforwardly, and it is given by
\begin{align}
    &\mathcal{P}_{\widetilde{f}^{(1,1)}}= \sum_{\blacksquare=1,U,V} \mathcal{P}_{\widetilde{f}^{(1,1)}_{\blacksquare}}=20\zeta(5)+20\zeta(5)+20\zeta(5)=60\zeta(5)\,.
\end{align}
In conclusion, we find
\begin{align}
I^{\prime}_4\left[{F^{(1)}}(x_i)\right]
= - 60\zeta(5) \, ,
\end{align}
which agrees with the localisation result given in \eqref{eq:d^4_m}.

At two loops, we have 
\begin{align}
I^{\prime}_4\left[{F^{(2)}}(x_i) \right] &= 4\times \frac{1}{2! (-4)^2} \times \left(\mathcal{P}_{\widetilde{f}^{(2,1)}}+\mathcal{P}_{\widetilde{f}^{(2,2)}}\right)\,,
\end{align}
where $\widetilde{f}^{(2,1)}$ and $\widetilde{f}^{(2,2)}$ are give by
\begin{align}
    &\widetilde{f}^{(2,1)}(x_i)=\sum_{\blacksquare=1, U, V}\widetilde{f}^{(2,1)}_{\blacksquare}(x_i)
    =\left(\frac{x_{13}^2x_{24}^2\,(1+U+V)}{\prod_{1\leq i < j\leq 4}x^2_{ij}}\, g \times h^{(5,6)}_{(1,2;3,4)}\right)+\mathcal{S}_4 \times \mathcal{S}_2 \nonumber \\
    &\widetilde{f}^{(2,2)}(x_i)=\sum_{\blacksquare =1, U, V}\widetilde{f}^{(2,2)}_{\blacksquare}(x_i)
    =\left(\frac{x_{13}^2x_{24}^2\,(1+U+V)}{\prod_{1\leq i < j\leq 4}x^2_{ij}}\, g \times \left[g^{(5)}_{(1,2,3,4)}\times g^{(6)}_{(1,2,3,4)} \right]\right)+\mathcal{S}_4\times \mathcal{S}_2 \,,
\end{align}
and $h^{(5,6)}_{(1,2;3,4)}$ is the integrand of a two loop ladder 
\begin{align}\label{eq:g+h}
 h^{(5,6)}_{(1,2;3,4)}& = \frac{x^2_{34}}{(x_{15}^2 x_{35}^2 x_{45}^2) x_{56}^2
(x_{26}^2 x_{36}^2 x_{46}^2)} \,.
\end{align}
Here $\mathcal{S}_4\times \mathcal{S}_L$ (for the case we are considering, $L=2$) means that we first sum over distinct permutation of $\mathcal{S}_4$ of the four external points $(x_1, x_2, x_3, x_4)$, and then sum over distinct permutations of $\mathcal{S}_L$ for the $L$ vertices that we integrate over. In practice, this implies
\begin{equation}
    (\cdots)+\mathcal{S}_4\times \mathcal{S}_L=\left((\cdots)+\mathcal{S}_4\right)+ \mathcal{S}_L=\left((\cdots)+\mathcal{S}_L\right)+ \mathcal{S}_4 \,.
\end{equation}
By `distinct permutation' we mean, for example, $h^{(5,6)}_{(1,2;3,4)}$ and $h^{(5,6)}_{(1,3;2,4)}$ are distinct under $\mathcal{S}_4$ permutation, while  $h^{(5,6)}_{(1,2;3,4)}$ and $h^{(5,6)}_{(2,1;3,4)}$ are not. Following such counting rules, we deduce $\widetilde{f}^{(2,1)}$ and $\widetilde{f}^{(2,2)}$ have $12$ and $3$ terms, respectively. We will again utilise the fact that the period for each term inside $\mathcal{S}_4\times\mathcal{S}_2$ permutations has the same value.
Their periods are explicitly given by
\begin{align}
    &\mathcal{P}_{\widetilde{f}^{(2,1)}}= \sum_{\blacksquare=1,U,V} \mathcal{P}_{\widetilde{f}^{(2,1)}_{\blacksquare}}=12\times\left(\frac{441}{8}\zeta(7)+70\zeta(7)+\frac{441}{8}\zeta(7)\right)=12\times\frac{721}{4}\zeta(7)\,, \\
    &\mathcal{P}_{\widetilde{f}^{(2,2)}}= \sum_{\blacksquare=1,U,V} \mathcal{P}_{\widetilde{f}^{(2,2)}_{\blacksquare}}=3\times\left[ 36\zeta(3)^2+\left(72\zeta(3)^2-21\zeta(7)\right)+36\zeta(3)^2\right]
    =3\times\left(144\zeta(3)^2-21\zeta(7)\right)\,. \nonumber
\end{align}
Using these results, we find
\begin{align}
I^{\prime}_4\left[{F^{(2)}}(x_i) \right] 
& =4\times \frac{1}{2! (-4)^2} \times\left(12\times \frac{721}{4}\zeta(7)+3 \times (144\zeta(3)^2-21\zeta(7))\right)\nonumber \\
&=\frac{3}{2} \times\left(36\zeta(3)^2+175\zeta(7)\right)\,,
\end{align}
which is in agreement with the result of supersymmetric localisation computation, as given in \eqref{eq:d^4_m}. 

At three loops, summing over $\widetilde{n}_3=5$ structures, we have 
\begin{align}
I^{\prime}_4\left[{F^{(3)}}(x_i)\right] &=4\times \frac{1}{3! (-4)^3} \times \left(\sum_{k=1}^5 \mathcal{P}_{\widetilde{f}^{(3,k)}}\right) \,,
\end{align}
where the ${\widetilde{f}^{(3,k)}}$ terms are give by (according to (4.16) of \cite{Eden:2012tu}), 
\begin{align} \label{eq:three_loop_tildef}
\widetilde{f}^{(3,1)}(x_i) = \left(\frac{x_{13}^2x_{24}^2\,(1+U+V)}{\prod_{1\leq i < j\leq 4}x^2_{ij}}\, g\times T^{(5,6,7)}_{(1,3;2,4)}\right)+\mathcal{S}_4\times \mathcal{S}_3 \, ,
\end{align}
\begin{align} 
\widetilde{f}^{(3,2)}(x_i) = \left(\frac{x_{13}^2x_{24}^2\,(1+U+V)}{\prod_{1\leq i < j\leq 4}x^2_{ij}}\, g\times E^{(5,6,7)}_{(1,3;2,4)}\right)+\mathcal{S}_4\times \mathcal{S}_3 \, ,
\end{align}
\begin{align} 
\widetilde{f}^{(3,3)}(x_i) = \left(\frac{x_{13}^2x_{24}^2\,(1+U+V)}{\prod_{1\leq i < j\leq 4}x^2_{ij}}\, g\times L^{(5,6,7)}_{(1,3;2,4)}\right)+\mathcal{S}_4\times \mathcal{S}_3\, ,
\end{align}
\begin{align} 
\widetilde{f}^{(3,4)}(x_i) = \left(\frac{x_{13}^2x_{24}^2\,(1+U+V)}{\prod_{1\leq i < j\leq 4}x^2_{ij}}\, g\times (g\times h)^{(5,6,7)}_{(1,3;2,4)}\right)+\mathcal{S}_4\times \mathcal{S}_3  \, ,
\end{align}
\begin{align} 
\widetilde{f}^{(3,5)}(x_i) = \left(\frac{x_{13}^2x_{24}^2\,(1+U+V)}{\prod_{1\leq i < j\leq 4}x^2_{ij}}\, g\times H^{(5,6,7)}_{(1,3;2,4)}\right)+\mathcal{S}_4\times \mathcal{S}_3 \,.
\end{align}
The explicit expressions of these integrands and their corresponding periods $\mathcal{P}_{\widetilde{f}^{(3,k)}}$ are given in appendix \ref{app3loops}. As an example, we have illustrated the Feynman graphs for $\widetilde{f}^{(3,5)}_1, \widetilde{f}^{(3,5)}_U, \widetilde{f}^{(3,5)}_V$ in Fig.\ref{fig:tilde_fgraphs_f3}. 
Using the results that are given in appendix \ref{app3loops} and after summing over all the periods $\mathcal{P}_{\widetilde{f}^{(3,k)}}$'s (in particular \eqref{eq:Ptf3}), we obtain, 
\begin{align}
I^{\prime}_4\left[{F^{(3)}}(x_i)\right] &=  4 \times \frac{1}{3! (-4)^3} \times \left(\sum_{k=1}^5 \mathcal{P}_{\widetilde{f}^{(3,k)}}\right)\nonumber \\
&=-\frac{45}{2}\times\left(20\zeta(3)\zeta(5)+49\zeta(9)\right)\,,
\end{align}
which again agrees with the localisation result (\ref{eq:d^4_m}) for the $a_{G_N}^3$ term.
\begin{figure} 
  \centering
  \begin{subfigure}{.3\linewidth}
    \centering\begin{tikzpicture}[scale=1.95, line width=1 pt]
    \begin{scope}[rotate=0]
       \coordinate (A1) at  (2.06, 0.1539);
       \coordinate (A2) at  (2.199, 0.8146);
       \coordinate (A3) at  (0.5106, 0.005049);
       \coordinate (A4) at  (0.5153, 1.28);
       \coordinate (A5) at  (1.162, 0.1818);
       \coordinate (A6) at  (-0.00112, 0.5261);
       \coordinate (A7) at  (1.055, 0.8703);
       \coordinate (A8) at  (1.497, 1.103);

       \draw (A1) to (A2);
       \draw (A2) to (A8);
       \draw (A8) to (A4);
       \draw (A4) to (A6);
       \draw (A6) to (A3);
       \draw (A3) to (A5);
       \draw (A5) to (A1);       
             \draw (A2) to (A5);
       \draw (A2) to (A7);
       \draw (A4) to (A7);
       \draw (A5) to (A6);
       \draw (A6) to (A7);

  \draw[line width=1mm,white] (A1) to (A7);
       \draw (A1) to (A7);       
       \draw[line width=1mm,white] (A3) to (A7);
       \draw (A3) to (A7);       
          \draw[line width=1mm,white] (A3) to (A8);
       \draw (A3) to (A8);       
          \draw[line width=1mm,white] (A1) to (A8);
       \draw (A1) to (A8);       
          \draw[line width=1mm,white] (A4) to (A5);
       \draw (A4) to (A5);       
       \draw[dashed,gray] (A7) to (A5);
       
\filldraw[black] (A1) circle (1.9pt) node[anchor=west]{};
     \filldraw[black] (A1) circle (1.5pt) node[anchor=west]{};
	\filldraw[black] (A2) circle (1.5pt) node[anchor=west]{};
	\filldraw[black] (A3) circle (1.5pt) node[anchor=west]{};
	\filldraw[black] (A4) circle (1.5pt) node[anchor=west]{};
	\filldraw[blue] (A5) circle (1.5pt) node[anchor=west]{};
    \filldraw[black] (A6) circle (1.5pt) node[anchor=west]{};
	\filldraw[blue] (A7) circle (1.5pt) node[anchor=west]{};
\filldraw[black] (A8) circle (1.75pt) node[anchor=west]{};
		\filldraw[gray] (A8) circle (1.5pt) node[anchor=west]{};
       \end{scope}
    \end{tikzpicture}
      \caption*{$\widetilde{f}^{(3,5)}_1$}
     \end{subfigure}
  \hspace{0.85em}
  \begin{subfigure}{.3\linewidth}
    \centering
    \begin{tikzpicture}[scale=2.15, line width=1 pt]
    \begin{scope}[rotate=0]
       \coordinate (A1) at  (0.3768,0.31 );
       \coordinate (A2) at  (0.3682, 1.521);
       \coordinate (A3) at  (1.142, 0.304);
       \coordinate (A4) at  (1.138, 1.525);
       \coordinate (A5) at  (1.322, 0.9144);
       \coordinate (A6) at  (2.032, 0.9144);
       \coordinate (A7) at  (0.7766, 0.9144);
      \coordinate (A8) at  (0.00277, 0.9144); 

       \draw (A1) to (A8);
       \draw (A2) to (A8);
       \draw (A2) to (A4);
       \draw (A4) to (A6);
       \draw (A3) to (A6);
       \draw (A3) to (A1);
      \draw (A3) to (A8);
       \draw (A4) to (A8);
      \draw (A3) to (A5);
       \draw (A4) to (A5);
      \draw (A5) to (A6);
    \draw (A3) to (A7);
       \draw (A4) to (A7);
       \draw[line width=1mm,white] (A1) to (A5);
       \draw (A1) to (A5);
       \draw[line width=1mm,white] (A2) to (A5);
       \draw (A2) to (A5);
          \draw[line width=1mm,white] (A1) to (A7);
       \draw (A1) to (A7);
       \draw[line width=1mm,white] (A2) to (A7);
       \draw (A2) to (A7);
       \draw[line width=1mm,white] (A6) to [bend left=15](A7);
\draw (A6) to [bend left=15](A7);
 \draw[dashed,gray] (A3) to (A4);
 \draw[dashed,gray] (A7) to (A5);
     
      \filldraw[black] (A1) circle (1.25pt) node[anchor=west]{};
	\filldraw[black] (A2) circle (1.25pt) node[anchor=west]{};
	\filldraw[blue] (A3) circle (1.25pt) node[anchor=west]{};
	\filldraw[blue] (A4) circle (1.25pt) node[anchor=west]{};
	\filldraw[blue] (A5) circle (1.25pt) node[anchor=west]{};
    \filldraw[black] (A6) circle (1.25pt) node[anchor=west]{};
	\filldraw[blue] (A7) circle (1.25pt) node[anchor=west]{};

\filldraw[black] (A8) circle (1.55pt) node[anchor=west]{};
\filldraw[gray] (A8) circle (1.25pt) node[anchor=west]{};
	\end{scope}
       \end{tikzpicture}
         \caption*{$\widetilde{f}^{(3,5)}_U$}
  \end{subfigure}
  \hspace{0.85em}
  \begin{subfigure}{.3\linewidth}
    \centering
   \begin{tikzpicture}[scale=1.85, line width=1 pt]
    \begin{scope}[rotate=0]
       \coordinate (A1) at  (1.699, 0.3725);
       \coordinate (A2) at  (1.71, 0.942);
       \coordinate (A3) at  (1.223, 1.342);
       \coordinate (A4) at  (1.223, -0.03177);
       \coordinate (A5) at  (0.751, 0.3621);
       \coordinate (A6) at  (-0.01082, 0.6368);
       \coordinate (A7) at  (0.7407, 0.9374);
 \coordinate (A8) at  (2.477, 0.6316);
    
     \draw (A3) to (A6);
     \draw (A6) to (A4);
     \draw (A4) to (A8);
\draw (A3) to (A8);
     \draw (A1) to (A2);
     \draw (A2) to (A7);
     \draw (A1) to (A5);
    \draw (A3) to (A7);
     \draw (A6) to (A7);
     \draw (A4) to (A5);
   \draw (A6) to (A5);
   \draw (A1) to (A8);
   \draw (A2) to (A8);
     \draw[line width=1mm,white] (A2) to (A5);
       \draw (A2) to (A5);
  \draw[line width=1mm,white] (A1) to (A7);
       \draw (A1) to (A7);
  \draw[line width=1mm,white] (A2) to (A4);
       \draw (A2) to (A4);
  \draw[line width=1mm,white] (A2) to (A7);
       \draw (A2) to (A7);
  \draw[line width=1mm,white] (A4) to (A7);
       \draw (A4) to (A7);
  \draw[line width=1mm,white] (A1) to (A3);
       \draw (A1) to (A3);
  \draw[line width=1mm,white] (A1) to (A5);
       \draw (A1) to (A5);
  \draw[line width=1mm,white] (A3) to (A5);
       \draw (A3) to (A5);
       \draw[dashed,gray] (A1) to (A4);
 \draw[dashed,gray] (A2) to (A3);
 \draw[dashed,gray] (A5) to (A7);
      
       \filldraw[blue] (A1) circle (1.35pt) node[anchor=west]{};
	\filldraw[blue] (A2) circle (1.35pt) node[anchor=west]{};
	\filldraw[blue] (A3) circle (1.35pt) node[anchor=west]{};
	\filldraw[blue] (A4) circle (1.35pt) node[anchor=west]{};
	\filldraw[blue] (A5) circle (1.35pt) node[anchor=west]{};
    \filldraw[black] (A6) circle (1.35pt) node[anchor=west]{};
	\filldraw[blue]  (A7) circle (1.35pt) node[anchor=west]{};
    \filldraw[black]  (A8) circle (1.65pt) node[anchor=west]{};
\filldraw[gray]  (A8) circle (1.35pt) node[anchor=west]{};
	\end{scope}
       \end{tikzpicture}
       \caption*{$\widetilde{f}^{(3,5)}_V$}
     \end{subfigure}
  \caption{Here we draw some examples of $\widetilde{f}$-graphs, $\widetilde{f}^{(3,5)}_1$,$\widetilde{f}^{(3,5)}_U$, $\widetilde{f}^{(3,5)}_V$, which are relevant for the three-loop computation. Once again, they are all non-planar diagrams.}
  \label{fig:tilde_fgraphs_f3}
\end{figure}
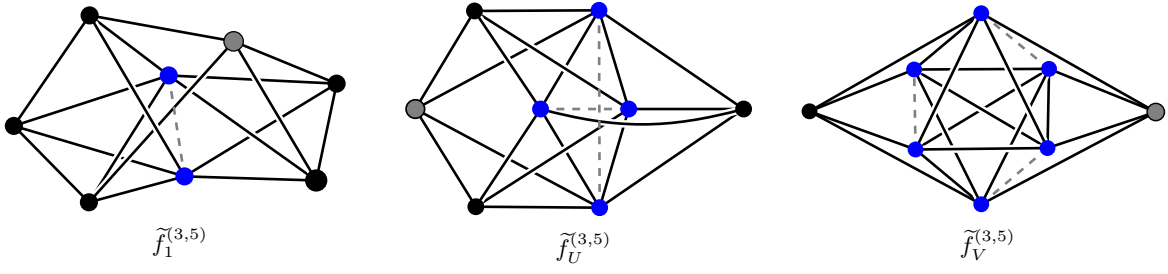
\section{Conclusion}
\label{sec:conclude}

In this paper, we studied the perturbative aspects of recently introduced integrated correlators in $\mathcal{N}=4$ SYM with general classical gauge groups. We identified interesting relations between the integrated correlators and periods of so-called $f$-graphs and their generalisation $\widetilde{f}$-graphs. $f$-graphs were used for constructing loop integrands of the un-integrated correlators. This identification paves the way of systematically computing integrated correlators to higher loops. In this paper, we applied the idea and computed all the relevant periods for the first integrated correlator up to four loops and for the second integrated correlator up to three loops, and we found the results perfectly agree with the expressions obtained from supersymmetric localisation. Our results extend the earlier computation of \cite{Dorigoni:2021guq}, where the first integrated correlator was computed for the first two loops. 

These explicit perturbative results verify the prediction of supersymmetrical localisation. Furthermore, the results also show explicitly the simplicity of the integrated correlators. A nice example of this is that even though the three-loop un-integrated correlator takes rather complicated form, the additional integration arising from the definition of integrated correlators simplify the structures drastically. It would be interesting to extend the computation of this paper to higher loops and the non-planar sectors, and systematically understand the simplicity of integrated correlators from Feynman diagram point of view, which may help to find new integration measures different from those considered in this paper that can also simply the un-integrated correlator. It will also be interesting to consider correlators involving higher-weight operators. In particular, the perturbative contribution of integrated four-point correlator of $\langle \mathcal{O}_2\mathcal{O}_2\mathcal{O}_p\mathcal{O}_p \rangle$ can also be computed using supersymmetric localisation \cite{Binder:2019jwn}. It would be interesting to analyse these integrated correlators with higher-weight operators using our methods and the corresponding integrands constructed in \cite{Chicherin:2015edu, Chicherin:2018avq} and more recently \cite{Caron-Huot:2021usw}. 

Our observation that relates the integrated correlators and periods also provides new non-trivial relations for the periods, given the fact that the integrated correlators can be computed using supersymmetric localisation exactly to any orders in the perturbation theory. This idea was illustrated and used to predict an analytical expression for a six-loop period, by using the localisation result of the first integrated correlator at five loops. It is of interest to verify our prediction by a direct computation of this particular six-loop period. It is also very interesting to understand these relations from mathematical viewpoints, and such an understanding will help to lead to systematic evaluation of the integrated correlators in the perturbation theory. In particular, each period may contain multiple zeta values, and their particular combinations to form integrated correlators, from the results of supersymmetric localisation, we know that these multiple zeta values should actually all cancel out when we add the periods together to form the integrated correlators. 

 \section*{Acknowledgements}
 
 We would like to thank Yu-tin Huang and Chia-Kai Kuo for the collaboration in the early stage of this project. We also thank Michael Borinsky,  Jacob Bourjaily, Gang Chen, Daniele Dorigoni, Michael Green, Paul Heslop, Erik Panzer, and Oliver Schnetz for useful conversations.   CW is supported by a Royal Society University Research Fellowship No. UF160350.  SZ is supported by a Royal Society grant RGF$\backslash$R1$\backslash$180037. 

\appendix
\section{Matrix model computations} \label{matrix_model}

In this appendix we review the perturbative contribution of the ${\cal N}=2^*$ SYM partition function on $S^4$,  $\hat Z^{pert}_{G_N}(m, a)$, for all classical gauge groups $G_N$, and the corresponding matrix model expectation values.\footnote{The non-perturbative instanton contributions to the partition function, the Nekrasov partition function \cite{Nekrasov:2002qd, Nekrasov:2003rj}, with general gauge groups can be found in \cite{Billo:2015pjb, Billo:2015jyt, Billo:2016zbf}, and their contributions to the first integrated correlator were studied in \cite{Dorigoni:2022zcr}.} The perturbative contribution for the first integrated correlator has been computed in \cite{Alday:2021vfb, Dorigoni:2022zcr}. Therefore, we use these expressions of $\hat Z^{pert}_{G_N}(m, a)$ mainly for the computations of the perturbative contribution of the second integrated correlators in the main text.  The expressions of $\hat Z^{pert}_{G_N}(m, a)$ for $G_N=SU(N), SO(2N), SO(2N+1), USp(2N)$ are listed below.

\begin{itemize}

 \item  For $SU(N)$, we have
\begin{align}
& \hat Z^{pert}_{SU(N)}(m, a) =   \frac{1}{H(m)^{N-1}}  \prod_{i<j}\frac{H^2(a_{ij})}{H(a_{ij}-m)H(a_{ij}+m)}\,, 
\end{align}
with $a_{ij}=a_i -a_j$, and the expectation value of a function $F(a_i)$ is defined as
\begin{align}
\langle F(a_i) \rangle_{SU(N)} = {1\over \mathcal{N}_{SU(N)} } \int d^N a\, \delta \left(\sum_{i} a_i \right) \left(\prod_{i<j} a_{ij}^2 \right) \,  e^{-\frac{8\pi^2}{g_{_{YM}}^2} \sum_i a^2_i }\,  F(a_i)\, ,
\end{align}
where $\mathcal{N}_{SU(N)}$ is a normalisation factor such that $\langle 1 \rangle_{SU(N)}=1$. The function $H(m)$ is defined as \begin{equation}
H(m)=e^{-(1+\gamma)m^2}\, G(1+im)\, G(1-im)\,,
 \end{equation}
 where $G(m)$ is a Barnes G-function (and $\gamma$ is the Euler constant). 

 \item For $SO(2N)$, we have
 \begin{align}
& \hat Z^{pert}_{SO(2N)}(m, a)=  \frac{1}{H(m)^N} \prod_{i<j}\frac{H^2(a_{ij})H^2(a_{ij}^+)}{H(a_{ij}-m)H(a_{ij}+m)H(a_{ij}^+-m)H(a_{ij}^++m)}\,, 
\end{align}
where $a_{ij}^+=a_i+a_j$,  and the expectation value is defined as
\begin{align}
\langle F(a_i) \rangle_{SO(2N)} = {1\over \mathcal{N}_{SO(2N)} } \int d^N a  \, \left (\prod_{i<j} a_{ij}^2 (a_{ij}^+)^2 \right)\,  e^{-\frac{8\pi^2}{g_{_{YM}}^2} \sum_i a^2_i}\,  F(a_i)\, .
\end{align}

 \item For $SO(2N+1)$, we have 
 \begin{align}
 & \hat Z^{pert}_{SO(2N+1)}(m,a) = \frac{1}{H(m)^N}\prod_i\frac{H^2(a_i)}{H(a_i+m)H(a_i-m)} \nonumber \\
 &\qquad \qquad \qquad \quad \times \prod_{i<j}\frac{H^2(a_{ij})H^2(a_{ij}^+)}{H(a_{ij}-m)H(a_{ij}+m)H(a_{ij}^+-m)H(a_{ij}^++m)}\,,
 \end{align}
 and the expectation value is defined as
\begin{align}
\langle F(a_i) \rangle_{SO(2N+1)} = {1\over \mathcal{N}_{SO(2N+1)} } \int d^N a  \, \left(\prod_{i} a_i^2 \right)\,\left (\prod_{i<j} a_{ij}^2 (a_{ij}^+)^2 \right)\,  e^{-\frac{8\pi^2}{g_{_{YM}}^2} \sum_i a^2_i }\,  F(a_i)\, .
\end{align}
When $N=1$ (i.e. for the correlator of $SO(3)$), one needs to rescale $g_{{_{YM}}}^2 \rightarrow 2\, g_{{_{YM}}}^2$ in the above formula, as discussed in \cite{Alday:2021vfb}.

  \item For $USp(2N)$, we have 
  \begin{align}
  & \hat Z^{pert}_{USp(2N)}(m, a)=  \frac{1}{H(m)^N}\prod_i\frac{H^2(2a_i)}{H(2a_i+m)H(2a_i-m)}
\nonumber\\
 &\qquad \qquad \qquad \quad \times \prod_{i<j}\frac{H^2(a_{ij})H^2(a_{ij}^+)}{H(a_{ij}-m)H(a_{ij}+m)H(a_{ij}^+-m)H(a_{ij}^++m)}\,, 
 \end{align}
  and the expectation value is defined below
\begin{align}
\langle F(a_i) \rangle_{USp(2N)} = {1\over \mathcal{N}_{USp(2N)} } \int d^N a  \, \left(\prod_{i} a_i^2 \right)\,\left (\prod_{i<j} a_{ij}^2 (a_{ij}^+)^2 \right)\,  e^{-\frac{16\pi^2}{g_{_{YM}}^2} \sum_i a^2_i}\,  F(a_i)\, .
\end{align}

\end{itemize}
\section{Periods for the second integrated correlator at three loops} \label{app3loops}
We list the relevant $f$-graphs and their periods for the three loop computations. The functions associated with $\widetilde{f}^{(3,k)}$'s in (\ref{eq:three_loop_tildef}) are given by
  \begin{align}
 T^{(5,6,7)}_{(1,2;3,4)}&= \frac{x_{34}^2   x_{17}^2 }{(x_{15}^2 x_{35}^2) (x_{16}^2  x_{46}^2) (x_{27}^2 x_{37}^2  x_{47}^2)
    (x_{56}^2 x_{57}^2 x_{67}^2)}\ ,\nonumber \\
E^{(5,6,7)}_{(1,2;3,4)} & = \frac{x^2_{23} x^2_{24} x^2_{16}}{(x_{15}^2 x_{25}^2 x_{35}^2)
x_{56}^2 (x_{26}^2 x_{36}^2 x^2_{46}) x^2_{67} (x_{17}^2 x_{27}^2 x_{47}^2)}
\, , \nonumber \\
L^{(5,6,7)}_{(1,2;3,4)} & = \frac{x^4_{34}}{(x_{15}^2 x_{35}^2 x_{45}^2) x_{56}^2
(x_{36}^2 x_{46}^2) x^2_{67} (x_{27}^2 x_{37}^2 x_{47}^2)} \, , \nonumber \\
  {({g\times h})}^{(5,6,7)}_{(1,2;3,4)} &= \frac{x_{12}^2 x_{34}^4}{(x_{15}^2  x_{25}^2  x_{35}^2 x_{45}^2) (x_{16}^2  x_{36}^2x_{46}^2) (x_{27}^2  x_{37}^2   x_{47}^2) x_{67}^2} \ ,\nonumber\\
H^{(5,6,7)}_{(1,2;3,4)} & =  \frac{x_{14}^2 x_{23}^2 x_{34}^2  x^2_{57}}{(x_{15}^2 x_{25}^2 x_{35}^2
x^2_{45}) x_{56}^2 (x_{36}^2 x^2_{46}) x^2_{67} (x_{17}^2 x_{27}^2 x^2_{37}
x_{47}^2)} \, .
\end{align}
The periods of $\widetilde{f}^{(3,k)}$'s  are given by
\begin{align} \label{eq:Ptf3}
\sum_{k=1}^5\mathcal{P}_{\widetilde{f}^{(3,k)}}= 2160\times \left[ 20 \zeta (3) \zeta (5)+49 \zeta (9) \right] \,,
\end{align}
where each term in the sum above is given by 
\begin{align} \label{eq:f3k}
 &\mathcal{P}_{\widetilde{f}^{(3,1)}}=72\times \left(16\zeta (3)^3+\frac{5402}{9} \zeta (9)\right) \, ,
\nonumber \\
&\mathcal{P}_{\widetilde{f}^{(3,2)}}=72\times \left(-48 \zeta (3)^3+240 \zeta (5) \zeta (3)+\frac{1231 \zeta (9)}{3}\right) \, ,\nonumber \\
 &\mathcal{P}_{\widetilde{f}^{(3,3)}}=36\times \left(16\zeta (3)^3+\frac{5402}{9} \zeta (9)\right)\, , 
\nonumber \\
 &\mathcal{P}_{\widetilde{f}^{(3,4)}}=36\times \left(48 \zeta (3)^3-144 \zeta (3)^2+432 \zeta (5) \zeta (3)+378 \zeta (7)-\frac{388 \zeta (9)}{3}\right)\, ,
\nonumber \\
&\mathcal{P}_{\widetilde{f}^{(3,5)}}= 36\times \left(144 \zeta (3)^2+288 \zeta (5) \zeta (3)-378 \zeta (7)+448 \zeta (9)\right) \,.
\end{align}
The factor $72$ and $36$ above are numbers of terms inside the $\mathcal{S}_4 \times \mathcal{S}_3$ permutations, and they have the same period value. We note that $\zeta (3)^3$ terms cancel out in the sum \eqref{eq:Ptf3}, and each  $\mathcal{P}_{\widetilde{f}^{(3,k)}}$ in \eqref{eq:f3k} consists of three periods
   $\mathcal{P}_{\widetilde{f}^{(3,k)}_\blacksquare}$'s with $\blacksquare=1,U,V$, therefore 
   \ie \label{eq:pf3k}
   \mathcal{P}_{\widetilde{f}^{(3,k)}} = \sum_{\blacksquare =1, U, V}\mathcal{P}_{\widetilde{f}^{(3,k)}_\blacksquare} \, ,
   \fe
and each $\mathcal{P}_{\widetilde{f}^{(3,k)}_\blacksquare}$ is listed below
\begin{align}
    &\mathcal{P}_{\widetilde{f}^{(3,1)}_1}=72\times\left(\frac{1567}{9}\zeta(9) + 8\zeta(3)^3\right) \nonumber\\
    &\mathcal{P}_{\widetilde{f}^{(3,1)}_U}=72\times\left(252\zeta(9)\right)  \nonumber\\
    &\mathcal{P}_{\widetilde{f}^{(3,1)}_V}=72\times\left(\frac{1567}{9}\zeta(9) + 8\zeta(3)^3\right) \,,
\end{align}
and
\begin{align}
    &\mathcal{P}_{\widetilde{f}^{(3,2)}_1}=72\times\left(120\zeta(3)\zeta(5) + \frac{727}{6}\zeta(9) - 24\zeta(3)^3\right)  \nonumber\\
    &\mathcal{P}_{\widetilde{f}^{(3,2)}_U}=72\times\left(168\zeta(9)\right)   \nonumber\\
    &\mathcal{P}_{\widetilde{f}^{(3,2)}_V}=72\times\left(120\zeta(3)\zeta(5) + \frac{727}{6}\zeta(9) - 24\zeta(3)^3\right)  \,,
\end{align}
and
\begin{align}
    &\mathcal{P}_{\widetilde{f}^{(3,3)}_1}=36\times\left(\frac{1567}{9}\zeta(9) + 8\zeta(3)^3\right) \nonumber\\
    &\mathcal{P}_{\widetilde{f}^{(3,3)}_U}=36\times\left(252\zeta(9)\right)  \nonumber\\
    &\mathcal{P}_{\widetilde{f}^{(3,3)}_V}=36\times\left(\frac{1567}{9}\zeta(9) + 8\zeta(3)^3\right) \,,
\end{align}
and
\begin{align}
    &\mathcal{P}_{\widetilde{f}^{(3,4)}_1}=36\times\left(-36\zeta(3)^2 + \frac{189}{2}\,\zeta(7) + 108\zeta(3)\zeta(5)\right) \nonumber\\
    &\mathcal{P}_{\widetilde{f}^{(3,4)}_U}=36\times\left(-72\zeta(3)^2 + 189\zeta(7) + 216\zeta(3)\,\zeta(5) + 48\zeta(3)^3 - \frac{388}{3}\zeta(9)\right)  \nonumber\\
    &\mathcal{P}_{\widetilde{f}^{(3,4)}_V}=36\times\left(-36\zeta(3)^2 + \frac{189}{2}\,\zeta(7) + 108\zeta(3)\zeta(5)\right)\,,
\end{align}
and
\begin{align}
    &\mathcal{P}_{\widetilde{f}^{(3,5)}_1}=36\times\left(120\zeta(3)\zeta(5) \right)\nonumber\\
    &\mathcal{P}_{\widetilde{f}^{(3,5)}_U}=36\times\left(24\zeta(3)\zeta(5) + \frac{2126}{9}\zeta(9) + 72\zeta(3)^2 - 189\zeta(7) + 16\zeta(3)^3 \right) \nonumber\\
    &\mathcal{P}_{\widetilde{f}^{(3,5)}_V}=36\times\left(144\zeta(3)\zeta(5) + \frac{1906}{9}\zeta(9) + 72\zeta(3)^2 - 189\zeta(7) - 16\zeta(3)^3\right) \,.
\end{align}
Using these results and \eqref{eq:pf3k}, one finds the expressions given in \eqref{eq:f3k}. 

\bibliographystyle{ssg}
\bibliography{period-ref}
	
\end{document}